\documentclass[11pt]{scrartcl}

\usepackage{graphicx} 
\usepackage{caption} 
\usepackage{subcaption} 
\usepackage[official]{eurosym} 
\usepackage{amsmath} 
\usepackage[printonlyused]{acronym} 
\usepackage{breqn} 
\usepackage{color}
\usepackage{float}
\usepackage{tablefootnote} 
\usepackage{threeparttable} 
\usepackage{comment} 
\usepackage{multirow}%
\usepackage{amsmath,amssymb,amsfonts}%
\usepackage{amsthm}%
\usepackage{mathrsfs}%
\usepackage[title]{appendix}%
\usepackage{xcolor}%
\usepackage{textcomp}%
\usepackage{manyfoot}%
\usepackage{booktabs}%
\usepackage{algorithm}%
\usepackage{algorithmicx}%
\usepackage{algpseudocode}%
\usepackage{listings}%
\usepackage{siunitx}%
\usepackage{mathtools}
\usepackage{lineno}

\usepackage{xfrac}
\usepackage{subcaption}
\usepackage{authblk}

\usepackage{natbib}
\begin{document}

\DeclareSIUnit\TWh{TWh}
\DeclareSIUnit\MWh{MWh}
\DeclareSIUnit\GWh{GWh}
\DeclareSIUnit\euros{EUR}

\newcommand{\punc}[1]{\,#1}
\newcommand{\pdb}[2]{\frac{\partial \, #1}{\partial \, #2}}
\newcommand{\eqnref}[2][]{Eq.~(\ref{eq:#2}#1)}

\newcommand{\ntot}{n_{\mathrm{tot}}}
\newcommand{\rtot}{r_{\mathrm{tot}}}
\newcommand{\Co}[1]{{\Gamma^{(#1)}}}
\newcommand{\la}[1]{{\lambda^{(#1)}}}
\newcommand{\sa}[1]{{\sigma^{(#1)}}}
\newcommand{\Lagrange}{\mathcal{L}}
\newcommand{\Rsub}[1]{R^{(#1)}}
\newcommand{\RRgz}{\mathbb{R}_{\geq0}}

\newcounter{mpFootnoteValueSaver}


\author[1]{Silvian M. Radke\thanks{silvian.radke@b-tu.de}}
\author[2]{Philipp C. Verpoort}
\author[3,2]{Falko Ueckerdt}
\author[1]{Felix Müsgens}

\affil[1]{Chair of Energy Economics, Brandenburg University of Technology Cottbus--Senftenberg, Cottbus, Germany}

\affil[2]{Energy Transition Lab, Potsdam Institute for Climate Impact Research, Potsdam, Germany}
\affil[3]{Interdisciplinary Transformation University Austria, Linz, Austria}

\title{Planning resilient hydrogen supply chains under disruption risk}
\date{May 2025}

\maketitle


\begin{abstract}
Despite growing concerns over energy security, infrastructure planning and modelling for emerging green fuel supply chains often neglect risks from supply disruptions. Using a stochastic optimisation model of EU hydrogen imports, we show that ‘naive’ infrastructure planning results in welfare losses of \SI{12}{\percent} (24 billion EUR) compared to risk-aware planning that anticipates supply disruptions. Despite requiring higher upfront investments, anticipatory planning achieves welfare levels close to those of an idealised system without disruptions, but entails a markedly different infrastructure configuration. Two complementary resilience strategies emerge: diversification across import corridors and strategic over-investment. This leads to increased intra-European transport capacity, a broader set of import pipelines, and investments in costly shipping terminals for hydrogen carriers. Our results show that incorporating supply risk considerations into infrastructure planning helps prevent the structural vulnerabilities seen in fossil fuel systems when designing future hydrogen supply chains.
\end{abstract}


\section*{Main}
In line with the RePowerEU plan \cite{RePowerEU}, several studies forecast a strong growth in hydrogen consumption in the European Union (EU). A substantial share of this hydrogen is expected to be imported because conditions for production are better in non-EU countries, especially in Northern Africa. These imports could be realised through ships or pipelines. Pipelines are the cheapest mode of transport for large volumes over short distances ($< \SI{2000}{\kilo\meter}$), especially when natural-gas infrastructure can be repurposed \cite{Niermann2021,ACER2021}. However, pipelines entail higher outage risks through accidents, sabotage, or political disagreement. In case of a pipeline outage, the economic impacts can be substantial, especially when a short-term switch to other trade routes or energy carriers cannot provide sufficient volumes. These risks should be accounted for when modelling hydrogen infrastructure, which makes the decision problem stochastic \cite{Lee2014}. 

Previous studies either neglected the risk of supply disruptions of green hydrogen or discussed them qualitatively. Many studies calculate the costs and availability of renewables and search for minimal levelised costs of hydrogen across countries \cite{Moritz2023, Sens2022_1, Sens2022_2, franzmann2023, benalcazar2024,Brandt2024, Genge2023, Genge2025, Genge2026,
Riera2023}. Other studies consider the uncertainty of the hydrogen ramp-up \cite{Odenweller2022,Egli2025} and uncertainties in hydrogen generation costs, technological development, and demand development \cite{Aldren2025, Kim2024},  but without endogenously analysing security of supply. A few studies address supply risks, but focus on the demand side \cite{Rezaee2024,Almansoori2012}  or employ robust optimisation (i.e., optimisation in case of outage) \cite{Caglayan2021, Yokoyama2018}. Others study mitigation of supply risks in fossil supply chains through local renewable energy sources \cite{Emenike2020, Lambert2022, DeRosa_2022} but without considering supply chain risks and resilience for green hydrogen.

Here, we conduct a cost-benefit analysis of diversifying hydrogen supply chains for EU imports in 2050. We build on methods by Refs.~\cite{Dantzig1955, Lee2014, Riepin2021} to develop a two-stage stochastic optimisation model that quantifies the effects of supply-chain risks on economic welfare with and without anticipation. We use this model to analyse the European future hydrogen market following the welfare maximisation presented in Ref. \cite{Antweiler2025}.

This approach allows us to demonstrate the overarching benefit of resilient infrastructure and the high cost of ignoring supply risks. For a medium level of risk, investments in resilient supply chains yield welfare gains of 12 percentage points. We show that an intra-European hydrogen pipeline system contributes to resilience to a similar degree as terminals for shipping-based hydrogen. Finally, we show that investments in back-up capacities are worthwhile even in the case of low risk.

\newpage

\section*{Welfare gains from anticipating supply risks}

\begin{figure}[H]
    \centering
    \includegraphics[width=1\linewidth]{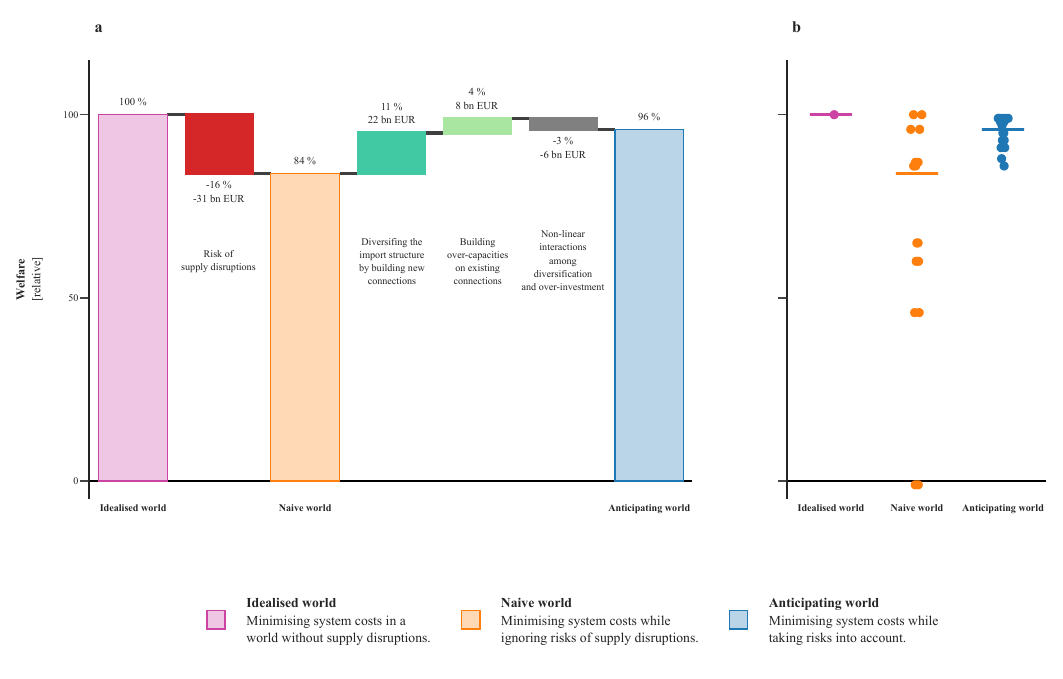}
    \caption{\textbf{Effects on welfare due to supply risks and anticipation.} We compare three scenarios: an \textit{idealised world} (theoretical benchmark: a world without risks), a \textit{naive world} (minimising costs while ignoring risks), and an \textit{anticipating world} (minimising costs while taking risks into account). \textbf{a}, Welfare is shown relative to the \textit{idealised world}. The bars for the \textit{naive} and \textit{anticipating worlds} represent expected welfare (which accounts for multiple realisations weighted by their respective probabilities). The waterfall diagram shows the attribution of the differences between these scenarios. First, the risk of supply disruptions causes a drop in the expected welfare in the \textit{naive world}. Second, diversification as one hedging strategy against this risk and, third, over-investment as another hedging strategy. Welfare gains are diminished by non-linear interactions between diversification and over-investment. \textbf{b}, We compare the variance in the welfare of all realisations (points) alongside expected welfare (lines) across scenarios.}
    \label{fig:results}
\end{figure}

Conventional energy system models `naively' optimise the cost of generation and transportation without anticipating supply risks. The resulting optimal infrastructure focusses purely on the cheapest options, as the model sees no benefits in resilience. However, several recent developments have shown that supply interruptions in commodity markets can occur: natural gas flows from Russia to Europe are one example, an ongoing discussion around rare earth exports from China another. At the same time, computational resources are ever improving, freeing computational power to model uncertainty endogenously. Here, we employ a stochastic model that anticipates risks and thereby attributes economic value to the resilience of infrastructure.

We model three distinct scenarios. First, an \textit{idealised world}, in which we abstract from supply risks. This is a theoretical benchmark, as reality entails risks. Consequently, risk is present in the other two scenarios, operationalised by allowing import pipeline connections to fail with non-zero probability (see Methods). In the second scenario, \textit{naive world}, the model ignores these risks in its investment decision. In other words, the optimisation falsely assumes that all suppliers deliver with a probability of \SI{100}{\percent}. This is the approach most common in the existing literature. The third scenario, an \textit{anticipating world}, takes into account the correct outage probabilities when investing. The resulting solution maximises welfare given the risk of supply disruptions.

We compare how economic welfare changes between these scenarios as risks and anticipation are accounted for (Fig.~\ref{fig:results}a). We find the expected welfare drops by as much as \SI{16}{\percent} in the \textit{naive} compared to the \textit{idealised world}, as the failing pipelines in the outage realisations lead to substantial supply shortages. Strikingly, if the model is allowed to account for these risks in the \textit{anticipating world}, welfare losses reduce from \SI{16}{\percent} to \SI{4}{\percent}.

Investors have two complementary options to adjust: diversification (reallocating capacities to include additional import corridors) and over-investment (increasing capacities on cheap risky import corridors and within Europe). We can quantify the individual importance of these two options by constraining our model to use only one of the two (see Methods). 11 percentage points of welfare loss are recovered through diversification and 4 percentage points through over-investment. In combination they yield 12 percentage points due to non-linear effects.

Our model of the European hydrogen supply network can choose to invest into 4 pipeline and 5 shipping connections for hydrogen imports and 4 pipeline connections for intra-European transport. We assume that outages for import pipelines are random and uncorrelated, with probabilities estimated based on country interest rates (see Methods and Fig.~\ref{fig:realisations}). The liquid hydrogen shipping and intra-European pipeline connections are assumed to be free of outage risk.

In addition to expected welfare (averaged over all realisations weighted by probabilities), we can also analyse the variation of welfare across all 16 realisations ($2^4$ combinations of outages) (Fig. \ref{fig:results}b). Welfare exhibits a huge spread in the \textit{naive world}, as realisations with no or very few outage realisations display high welfare, while others, with several vital connections failing, exhibit large welfare losses. In contrast, the variation of welfare in the \textit{anticipating world} is much lower. Consequently, the standard deviation of welfare decreases from 63 billion EUR in the \textit{naive world} to 8 billion EUR in the \textit{anticipating world} scenario. Note that the best-performing realisation is the one without pipeline outages and that its welfare in the \textit{naive world} is equal to that in the \textit{idealised world}, whereas its welfare is lower in the \textit{anticipating world} due to higher investment costs for infrastructure. However, these higher costs are overcompensated by the benefits of resilience in the resulting expected welfare. 

\begin{figure}[H]
    \begin{subfigure}{1\textwidth}
        \centering
        \includegraphics[width=1\linewidth]{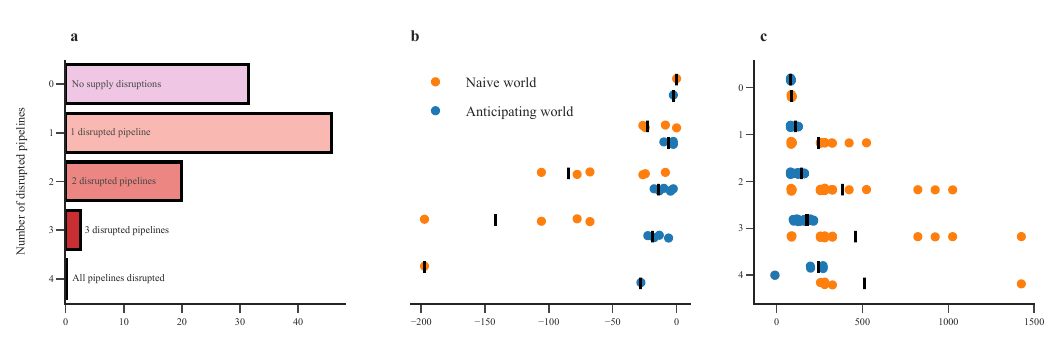}
        \label{fig:fig2_reals}
    \end{subfigure}
    \caption{\textbf{Probabilities, welfare losses, and hydrogen prices of realisations grouped by number of failed pipelines.} We show results for 2 scenarios (\textit{naive} and \textit{anticipating world}) and for 5 groups of realisations (16 realisations grouped by number of failed pipelines ranging from 0 to 4). All y-axes show the number of outages ranging from 0 to 4 (top to bottom). \textbf{a}, Probability of each group (i.e. combined probability of all realisations in the respective group). \textbf{b}, Distribution of welfare losses compared to the \textit{idealised world}. Each dot represents one realisation within the group and the black bars states the conditional expected value. \textbf{c}, distribution of hydrogen prices in each group. Each dot represents one demand node in one realisation. The bars present the conditional expected value. All nodes are weighted equal.}
    \label{fig:fig2}
\end{figure}

We also group the realisations by number of failed pipelines and analyse the probabilities and the distribution of economic welfare and hydrogen prices of each group in the \textit{naive} and \textit{anticipating worlds} (Fig. \ref{fig:fig2}). Realisations with 1 or 2 failed pipelines occur at a high combined probability of, respectively, \SI{45}{\percent} or \SI{20}{\percent} and exhibit high welfare losses and hydrogen prices in the \textit{naive} and moderate losses and prices in the \textit{anticipating world}. In contrast, realisations with 3 or 4 failed pipelines occur at a low probability of \SI{2.5}{\percent} or \SI{0.1}{\percent}, but exhibit high welfare losses and prices in the \textit{anticipating} and tremendously high in the \textit{naive world}. Groups with high welfare losses and hydrogen prices in average also show a huge variance, which indicates that the high losses and prices are strongly driven by a few realisations.

\begin{figure}[H]
    \centering
    \includegraphics[width=1\linewidth]{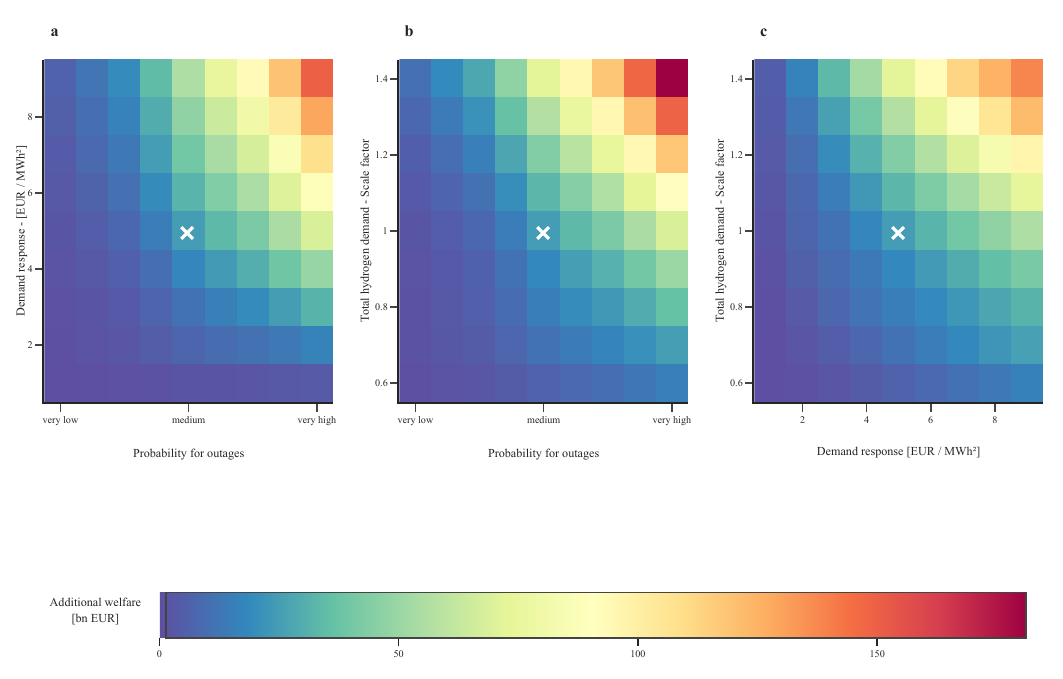}
    \caption{\textbf{Sensitivity analysis.} \textbf{a-c}, Heat maps represent welfare gains of a resilient supply chain. Welfare gains are calculated as the welfare difference between the \textit{anticipating} and \textit{naive worlds}. For each heat map one parameter (total hydrogen demand in a, demand response in b and probability of outage in c) is fixed to the value used in the empirical study. The outage risks across pipeline connections are presented in Table~\ref{tab:probabilities}. The white cross marks the parameters used above (Figs. 1 and 2).}
    \label{fig:sensitivity}
\end{figure}

The probability of pipeline outages, the total hydrogen import demand, and the demand response are tested, as these parameters strongly influence results but are particularly hard to estimate. We perform sensitivity analysis by varying these parameter across wide ranges when optimising our model, which confirms their strong influence on our results (Fig. \ref{fig:sensitivity}). This analysis also shows that there are no local optima, and extreme assumptions are needed to get extreme results. Especially the probability for outages is tested to extraordinary high values (over \SI{60}{\percent} for the pipelines from Northern Africa).

\section*{Resilient supply chain configuration}
\label{sec:networks}

\begin{figure}[H]
    \begin{subfigure}{1\textwidth}
        \centering
        \includegraphics[width=1\linewidth]{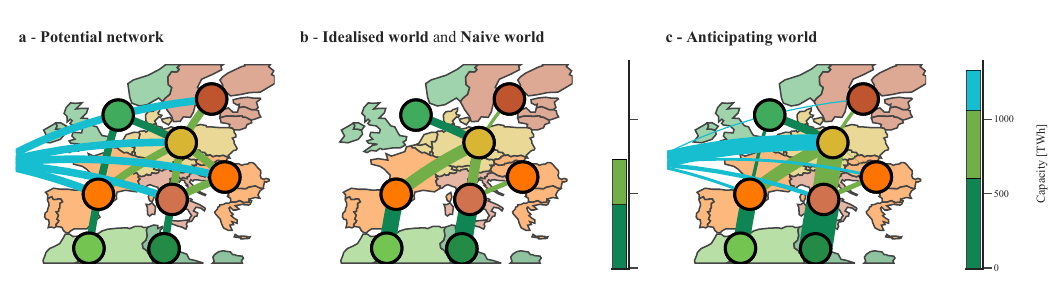}
        \label{fig:capacities}
    \end{subfigure}
    \begin{subfigure}{.9\textwidth}
        \centering
        \includegraphics[width=1\linewidth]{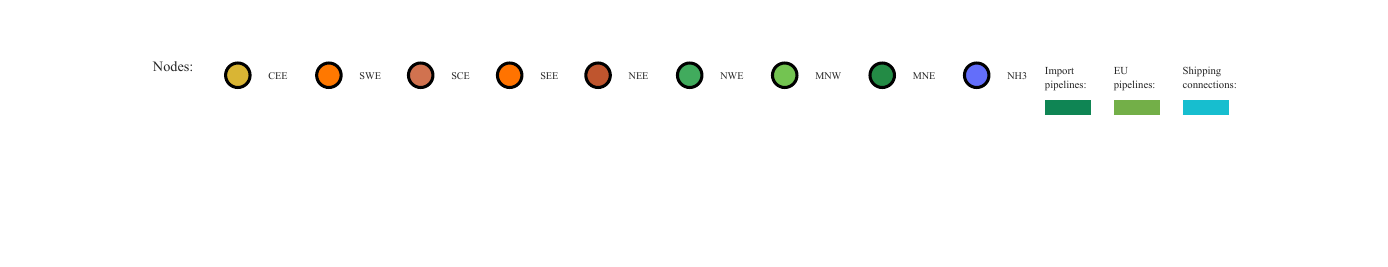}
        \label{fig:map_legend}
    \end{subfigure}
    \caption{\textbf{Type, amount, and geographic structure of optimal investment across scenarios.} \textbf{a}, This panel illustrates the model (which investments can be made) not its solutions (how investments are made in a specific scenario). The network consists of 9 nodes and 14 edges. 5 European demand nodes are shown in orange/brown, 3 supply nodes exporting via pipeline in green, and 1 supply node for shipping in blue. Pipelines (green edges) differentiate between import pipelines to Europe (dark green) and pipelines within Europe (light green). Shipping can reach all demand nodes (blue edges). The colouring indicates which node the countries in the map belong to. \textbf{b}, Investments in the \textit{idealised} and \textit{naive worlds}. \textbf{c}, Investments in the \textit{anticipating world}. The width of edges represents the amount of the transport capacity. The stacked bars beside the maps illustrate the aggregated terminal and pipeline capacities. }
    \label{fig:maps}
\end{figure}

We analyse the implications of risks and anticipation on investment in infrastructure regarding type and amount of capacity built (Fig. \ref{fig:maps}b and c). The realised capacities in the \textit{idealised} and \textit{naive worlds} are the same by design and facilitate the cheapest supply while risks are absent or ignored. As a result of the optimisation, hydrogen is imported by pipeline from North Africa and Norway (Fig. \ref{fig:maps}b). Pipeline investment within the EU is limited to what is strictly needed to deliver these imports to their final destination (e.g. from Northern Africa via Spain or Italy to Central Europe). No shipping terminals are built. In contrast, the \textit{anticipating world} builds terminals for more expensive imports via ship, especially in central Europe, and back-up pipeline capacities to North Africa and within Europe for additional flexibility (Fig. \ref{fig:maps}c). Consequently, the aggregated annual capacities nearly double from \SI{727}{\TWh} annual transmission capacity in the \textit{naive world} to \SI{1332}{\TWh} (\SI{1060}{\TWh} pipeline and \SI{272}{\TWh} terminal) capacity in the \textit{anticipating world}. This is also accompanied by an approximately 2.5-fold increase in investment costs from (\SI{1.75}{bn~EUR} in the \textit{naive world} to \SI{4.31}{bn~EUR} in the \textit{anticipating world}). These costs are initially significantly lower than assumed in the literature~\cite{Kotek2023,Kountouris2024}. However, this is put into perspective when one considers that, due to the clustering, our model examines a greatly simplified network and does not take distribution into account.

\section*{Analytical solution of reduced model}

\begin{figure}[H]
    \begin{subfigure}{1\textwidth}
        \centering
        \includegraphics[width=1\linewidth]{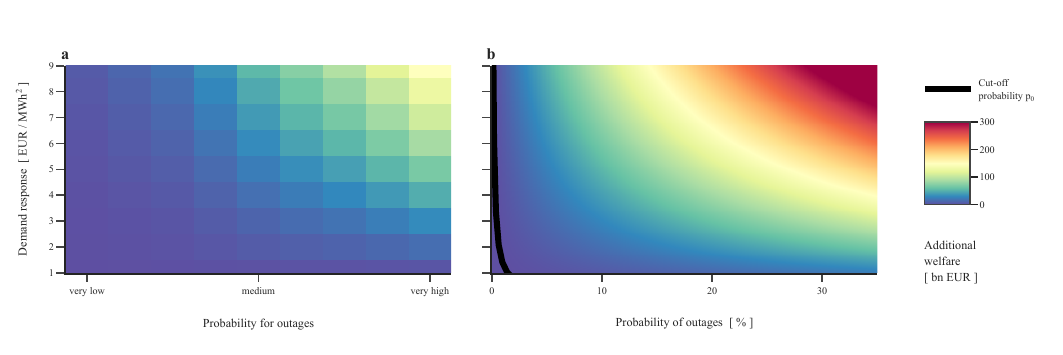}
        \label{fig:absolute_heatmap}
    \end{subfigure}
    \begin{subfigure}{1\textwidth}
        \centering
        \includegraphics[width=1\linewidth]{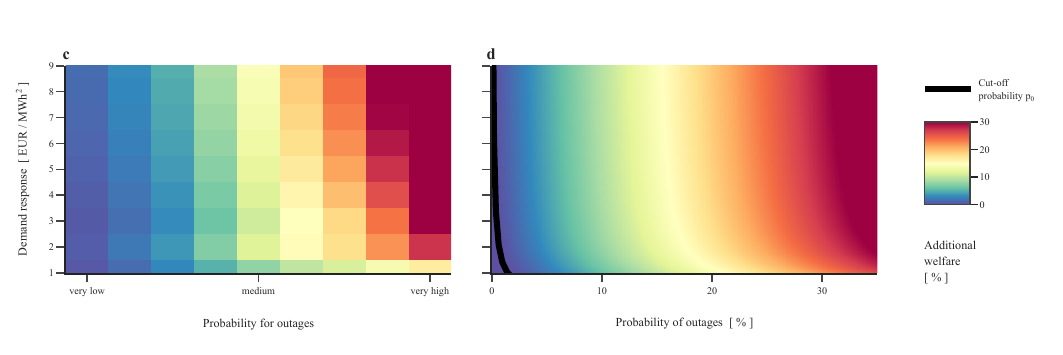}
        \label{fig:relative_heatmap}
    \end{subfigure}
    \caption{\textbf{Comparing welfare gains of full and reduced model across parameter ranges.} \textbf{a \& b}, The absolute welfare gain is calculated as the welfare of the \textit{anticipating world} reduced by the welfare of the \textit{naive world}. \textbf{c~\&~d}, the relative welfare gain is calculated as the \textit{anticipating world's} welfare reduced by the \textit{naive world's} normalised by the \textit{idealised world's} welfare. For the heat map in a the total demand is fixed to the level used in the empirical study while the demand response varies. The outage risks across pipeline connections are presented in Table~\ref{tab:probabilities}.}
    \label{fig:heatmap}
\end{figure}

To complement our empirical analysis, we analytically solve a reduced model that consists of only one demand node and two supply nodes. We find that the reduced model yields results for welfare that are similar to those of the full model (Fig.~\ref{fig:heatmap}). Consequently, we use the reduced model for deeper insights on the relative importance of key parameters. 

In addition to the absolute welfare gains of a more resilient supply chain, calculated as the difference between welfare $W$ in the \textit{anticipating world} and \textit{naive world} scenario $W_{\mathrm{Anticipating}}-W_{\mathrm{Naive}}$ (Fig. \ref{fig:heatmap} a \& b), we now also consider the relative welfare gains ${\left(W_{\mathrm{Anticipating}}-W_{\mathrm{Naive}}\right)}/{W_{\mathrm{Idealised}}}$ (Fig. \ref{fig:heatmap} c \& d), where the \textit{idealised world} acts as benchmark. These two metrics correspond to the \SI{24}{bn~EUR} and \SI{12}{\percent} welfare gains discussed above (Fig. \ref{fig:results}). The two metrics emphasise that the welfare gains are strongly driven by the maximal possible welfare, which in turn strongly depends on the demand response and the total hydrogen demand.

First, we find that the optimal solution contains no shipping terminals if the outage probability $p$ is below the threshold
\begin{equation}
    p^* = \frac{Fixed~Cost~Shipping}{Demand~Response\times Total~Demand - Variable~Cost~Shipping} \punc.
\end{equation}
No shipping terminals are built when welfare gains (denominator) do not outweigh the cost of constructing the terminal capacity for shipping (numerator) (see black line in \ref{fig:heatmap}b). For parameters close to those assumed in Fig. \ref{fig:results}, we find $p^* = \SI{2}{\percent}$. This finding suggests that investments that anticipate supply risks make sense even for relatively low outage risks.

Second, the analytical approach can also be used to quantify the value of risk-aware planning, i.e. the difference in welfare between the \textit{anticipating world} and the \textit{naive world} scenario. Assuming that $p>p^*$ and that the welfare of hydrogen consumption are large compared to the costs of importing hydrogen, the analytical calculation yields
\begin{equation}
      W_{\mathrm{Anticipating}}-W_{\mathrm{Naive}} \approx p \times W_{\mathrm{Idealised}} \times \left(1-\frac{Supply\text{-}cost~Premium}{Maximal~Welfare}\right) \punc,
\end{equation}
where the supply-cost premium is given as the difference between the pipeline-based and shipping-based hydrogen supply costs and the maximum welfare is the total welfare if hydrogen could be supplied for free. Therefore, the relative welfare gains are, to first-order approximation, given simply by the outage probability $p$, and, to second order, diminished by the supply-cost premium. This finding underlines the importance of the parameters $p$ and the magnitude of the total welfare.

\section*{Discussion and conclusions}
\label{sec:discussion}
Decades of supply shocks have taught industrialised countries a costly lesson in vulnerability. Two oil crises in the 1970s were followed by a natural gas supply crisis in 2022, when Europe’s reliance on imports from Russia was exploited during the Russian invasion of Ukraine, and by the Iran conflict in 2026, which severely affected the global economy — both within the energy sector and beyond. 
The “weaponisation” of energy supply can exploit existing vulnerabilities in geopolitical conflicts even without physical interruptions, for example through credible threats of supply disruption. Furthermore, the impacts of energy supply shocks can extend beyond economics and lead to social and political turbulence. Resilience has therefore emerged as a central priority in both public discourse and academic research.

Against the backdrop of heightened geopolitical tensions and a more contested international environment, the lesson becomes even clearer: an economy should not be too reliant on any specific energy supplier.
Consequently, the European Green Deal takes into account the risk of concentrating supply among too few providers and sets out the objective of diversifying energy supply chains \cite{RePowerEU, DeRosa_2022}.
This emphasises the political significance of risk-aware planning that accounts for disturbances and increases resilience.

Optimal planning that takes into account the risk of supply interruptions will consider diversification as well as strategic over-investments as important mitigation strategies. However, as infrastructure is costly, an optimum needs to be found between too little investment (low investment cost but severe damage from supply interruptions) and too much investment (high investment cost but low damage from supply interruptions).

Our work presents a novel two-stage stochastic optimisation model of EU hydrogen imports that highlights the advantages of risk-aware planning and efficient preparation for disruptions. The economic analysis yields welfare gains of about \SI{24}{bn~EUR} (\SI{12}{\percent}) relative to `naive' planning, i.e. ignoring the possibility of disruptions. Strikingly, this anticipatory approach achieves welfare levels close to those in an \textit{idealised world} without disruptions, but it requires a markedly different configuration of import routes and infrastructure.

Risk-aware planning applies two complementary hedging strategies: diversification across import corridors and over-investment to create back-up capacities. Consequently, hydrogen import infrastructure under risk-awareness features increased intra-European transport capacity, a more diverse set of import pipelines connecting nearby suppliers to the EU, and terminal capacities for shipping-based imports of liquid hydrogen carriers.
Note that our model chooses to invest in terminal capacities for shipping despite their increased cost compared to pipelines, which emphasises the role that terminal capacities play in resilience.
While not used under normal system conditions, the advantage of resilient infrastructure during rare supply interruptions outweighs its additional costs. If instead risks are ignored or very small (outage probability less than \SI{2}{\percent} in our parametrisation), the infrastructure safety-net appears to be `over-investments' or `stranded assets'.

Using sensitivity analysis and an analytical model, we provide deeper insight into the underlying mechanisms of resilient planning. Investment in resilience is optimal down to very low failure probabilities and that welfare gains are primarily dependent on the size of the market for green hydrogen (determined by latent demand and the demand-price response) and the probability of outage. While we have estimated these crucial parameters, they remain highly uncertain, which future work could address.

Applying our two-stage stochastic approach to infrastructure expansion models with a higher temporal and geographical resolution would allow for further insights into the hydrogen transport network and the role of hydrogen storage in the context of supply chain security.
Especially a higher geographical resolution including a less pronounced clustering appears promising. Based on our clustered network, the pipeline network within Europe is underestimated.
Our modelling optimises for expected welfare (weighted only by outage probability). Accounting for risk aversion (stronger weighting of realisations with high losses) would yield even more pronounced results, which could be analysed in future work.

Future infrastructure planning for emerging green fuels should avoid repeating the missteps observed in fossil fuel supply systems. Moving beyond short-term debates on securing fossil fuel supply, our results highlight the importance of long-term public investment in resilient green infrastructure. However, private actors may not take this into account in their planning. Political intervention may be necessary to overcome this shortcoming.
Policymakers may want to consider the option to require infrastructure planning to account for higher levels of resilience. As our empirical results show a significant decrease in welfare in cases with more than one disrupted pipeline, the conclusion of planners could be to increase the stability requirements to allow two relevant infrastructure items to fail without supply interruptions.
The increased cost could either be covered by hydrogen consumers or by the general public through subsidies, a decision that must be taken politically and should consider European industrial competitiveness \cite{Verpoort2024}, and a general need to stabilise the hydrogen ramp-up \cite{Odenweller2025} through lower cost and reduced uncertainty.

Future infrastructure planning for emerging green fuels should avoid repeating the missteps of fossil fuel infrastructure. Moving beyond current debates focused on compensation and securing fossil fuel supply, we hope our work will catalyse a broader, long-term discussion on public investment in resilient green infrastructure.

\newpage
\section*{Methods}
\label{sec:meth}

\subsection*{Model description}
\label{subsec:ESM}

Stochastic programming is a well-established tool used in energy systems modelling to address uncertainties \cite{herbst2012,pfenninger2014}. Based on the foundations laid by \cite{Dantzig1955} in 1955, state of the art models are used today to answer empirical questions in energy systems, see e.g. \cite{Lee2014,Riepin2019,Riepin2021,bernecker2022}. This section describes the two-stage stochastic optimisation model and the parametrisation chosen for our empirical study. 

\subsubsection*{Model algebra}
\label{subsubsec:model}
Sets, variables, and parameters of our model are defined in Tab.~\ref{tab:formula_characters}. Parameters, variables, and equations are defined over nodes $\{n\}$, edges $\{(n, n')\}$, and realisations $\{r\}$. Nodes correspond to locations of hydrogen demand and production, edges correspond to optional transportation corridors of hydrogen between nodes and uncertainty, realisations correspond to different outage cases of the import pipelines.

Our model maximises economic welfare, which is defined as the sum of consumer and producer rents. This is equivalent to minimising the costs in the second stage, defined as the sum of production costs, transport costs, and consumer rent losses. This equivalence holds because the sum of the two is the gross welfare, which is entirely determined by exogenous parameters. We also take into account the investment costs (first stage of the model), which are added to the costs of the second stage. Consequently, we minimise
\begin{align}
    TC = &\underbrace{\sum \limits_{n,n'}  K_{n,n'} \cdot c^f_{n,n'}}_{\textrm{infrastructure investment cost}} + \underbrace{\sum\limits_{r} p^r \sum\limits_{n} \Gamma^r_{n} \cdot c^{g}_{n}}_{\textrm{generation cost}}\notag\\ + &\underbrace{\sum\limits_{r} p^r \sum\limits_{n,n'} \Phi^r_{n,n'} \cdot c^{t}_{n,n'}}_{\textrm{transport variable cost}} + \underbrace{\sum\limits_{r} p^r \sum\limits_{n,n'}\left[ d_{n} - \Upsilon^r_{n}
    \right]^2 \cdot \frac{m_{n}}{2}}_{\textrm{consumer rent losses}}
\end{align}
\label{eq:objective}
The model is constrained by a continuity condition (consumption plus exports must equal production plus imports),
\begin{equation}
    \Gamma^r_{n} - \Upsilon^r_{n} + \sum\limits_{n'} (\Phi^r_{n',n} - \Phi^r_{n,n'} ) = 0 \punc,
    \label{eq:demand}
\end{equation}
and a capacity condition (the amount transported via an edge must be less or equal than the available capacity), 
\begin{equation}
    \Phi^r_{n,n'} \leq K_{n,n'} \cdot af^r_{n,n'} \punc.
\end{equation}
Moreover, generation and transport capacities are constrained by feasibility limitations,
\begin{equation}
    K_{n,n'} \leq \kappa_{n,n'} ~~ \text{and} ~~ \Gamma^r_{n} \leq \gamma_{n} \punc.
\end{equation}

The model setup is entirely green field, such that pipeline and terminal capacities have to be built up from zero. The model comprises one time step that represents one year (2050).

\begin{figure}[H]
    \centering
    \includegraphics[width=.9\linewidth]{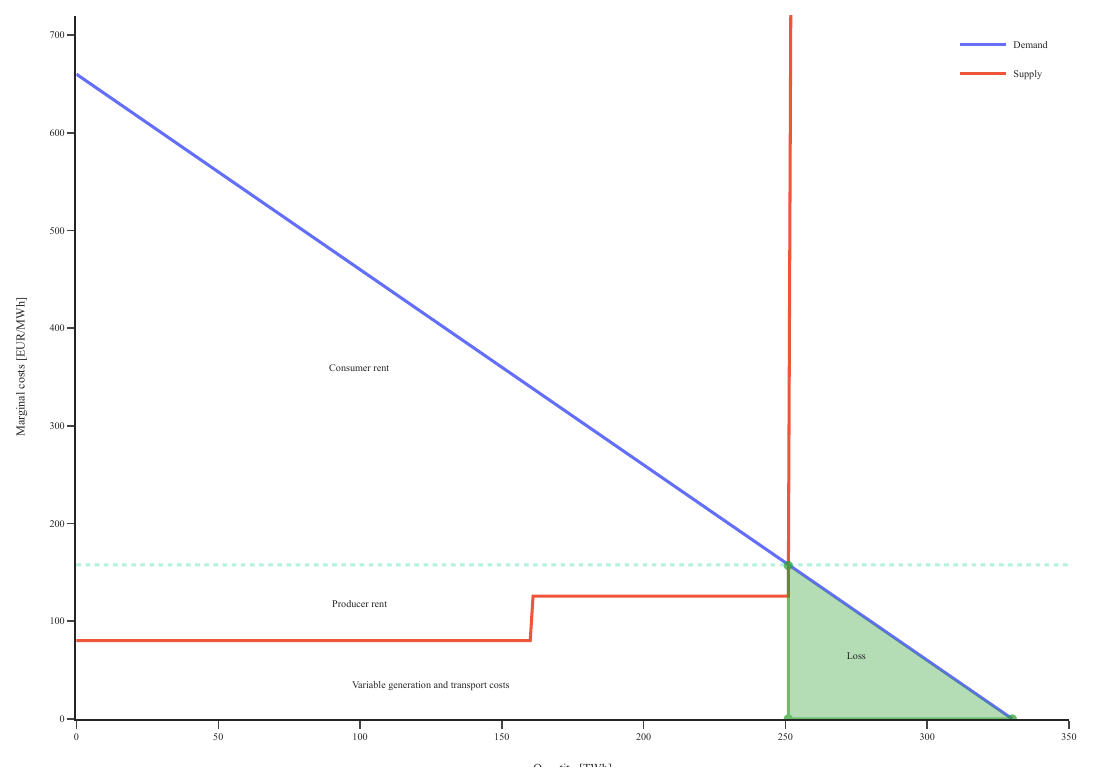}
    \caption{\textbf{Methodical approach to maximising welfare.} From an economic perspective, consumer and producer surplus are maximised. We achieve this by subtracting variable generation and transport costs ($c^g$, $c^t$) and loss (in green) from maximum welfare (the entire triangle under the blue demand function).}
    \label{fig:method}
\end{figure}

The optimisation takes place in two stages: first, investment into pipelines and terminals based on all possible realisations and, second, dispatch of generation, transport, and consumption given a specific realisation. The investment costs for transportation infrastructure in the first stage of the stochastic problem are independent of the uncertainty realisation $r$. The costs in the second stage of the problem include the costs to produce the desired amount of hydrogen, the variable transportation costs along the edges, and the consumer rent losses when supplying a node with less than latent demand (demand at price of zero). We follow methodology developed by Refs.~\cite{DeJonghe2012,Antweiler2021,Antweiler2025} and use a linear demand-response curve, which yields a quadratic term for consumer rent losses. For simplicity, we assume that fixed costs for generation capacity in a node are part of the variable production costs at a node. Fig.~\ref{fig:method} visualises the second stage of the welfare maximising approach.

\subsubsection*{Nomenclature}
\label{subsubsec:nomen}

Table~\ref{tab:formula_characters} shows the notation used in this paper, as well as the deviating notation used in GAMS.

\begin{threeparttable}
\caption{Notation}
\small
\label{tab:formula_characters}%
\setcounter{mpFootnoteValueSaver}{\value{footnote}} 

\begin{tabular}{llp{4cm}cc} 
\toprule
\multicolumn{2}{l}{Formula character}       & Description          & Unit           & Domain\\ 
General & GAMS                              &                      &                &       \\  \midrule
        &                                   &                      &                &       \\
\multicolumn{5}{l}{Sets}\\ \midrule
$r$     & $r$                       & Uncertainty realisation      &   -            &       \\
$n$, $n'$   & $n,nn$                & Nodes (and alias for nodes)  &   -            &       \\
            &                       &                              &                &       \\
\multicolumn{2}{l}{Parameters}      &                              &                &       \\ \midrule 
\multicolumn{5}{l}{Demand function}\\
\multicolumn{5}{l}{ }\\
$d_{n}$       & $dem(n)$ & Latent hydrogen demand in node $n$ (maximum import demand if the price is zero)  & \si{\TWh}    & $\mathbb{R}^{\geq0}$ \\

$m_{n}$ & $m(n)$           & Demand price response  
& Million \si{\euros\per\TWh\squared}   & $\mathbb{R}^{\geq0}$ \\
\multicolumn{5}{l}{ }\\
\multicolumn{5}{l}{Costs}\\
\multicolumn{5}{l}{ }\\
$c^{f}_{n,n'}$ & $fc(n,nn)$ & Fixed costs for transport capacities along edges & Million~\si{\euros\per\TWh} & $\mathbb{R}^{\geq0}$ 
\\
$c^g_{n}$ & $vc\_g(n)$ & Variable generation costs of node $n$      & Million~\si{\euros\per\TWh}       & $\mathbb{R}^{\geq0}$ \\
$c^t_{n,n'}$ & $vc\_t(n,nn)$ & Variable transport costs of edge $(n,n')$     & Million~\si{\euros\per\TWh}     & $\mathbb{R}^{\geq0}$   \\
\multicolumn{5}{l}{ }\\
\multicolumn{5}{l}{Limits}\\
\multicolumn{5}{l}{ }\\
$\gamma_n$ & $gen\_m(n)$ & Maximum hydrogen generation of node $n$    & \si{\TWh}   & $\mathbb{R}^{\geq0}$          \\
$\kappa_{n,n'}$ & $cap\_m(n,nn)$ & Maximum transportation capacity of edge $(n,n')$    & \si{\TWh}     
& $\mathbb{R}^{\geq0}$ \\
\multicolumn{5}{l}{ }\\
\multicolumn{5}{l}{Uncertainty}\\
\multicolumn{5}{l}{ }\\
$af_{n,n'}^r$  & $af(n,nn,r)$ & Availability factor of edge $(n,n')$     &  -  & $\{0,1\}$            \\
$p^r$        & $prob(r)$ & Probability for uncertainty realisation $r$&  -       &$[0,1]$     \\
                                  &                                    &       &     \\
\multicolumn{2}{l}{Variables}     &                                    &        &    \\ \midrule 
$\Upsilon_{n}^r$         & $USE(n,r)$ & Consumption of hydrogen in node $n$  & \si{\TWh} & $\mathbb{R}^{\geq0}$ \\
$\Gamma_{n}^r$    & $GEN(n,r)$ & Hydrogen generation in node $n$      & \si{\TWh} & $\mathbb{R}^{\geq0}$ \\
$K_{n,n'}$     & $CAP(n,nn)$ & Transport capacity of an edge     & \si{\TWh}        
& $\mathbb{R}^{\geq0}$ \\
$\Phi_{n,n'}^r$      & $FLOW(n,nn,r)$ & Flows between nodes            & \si{\TWh} & $\mathbb{R}^{\geq0}$ \\
                  &                &                                &            &\\ \bottomrule
\end{tabular}
\end{threeparttable}

\newpage

\subsubsection*{Restricted model for analysing strategies}
\label{subsubsec:restricting_strategies}

As discussed in the main text, two strategies could reduce welfare losses in case of supply outage: diversification (redistributing some of the infrastructure investments to other import routes) and over-investment (increasing the size of planned infrastructure). While the model can freely choose to employ both strategies in the \textit{anticipating world} scenario, we perform optimisations under additional constraints to attribute the welfare gains to one of these two strategies (Fig.~\ref{fig:results}).

Restricting the model to only diversification is achieved by adding a global limit to the capacity investments on all edges,
\begin{equation}
    \sum\limits_{n,n'}  K_{n,n'} = \sum\limits_{n,n'}  K_{n,n'}^{\textrm{Naive}}
    \label{eq:cap_max_div}
\end{equation}
where $K_{n,n'}^{\textrm{Naive}}$ are the capacity investments in the \textit{naive world}. Restricting the model to only over-investment is achieved by adding a constraint that forces $K_{n,n'} = 0$ if $K_{n,n'}^{\textrm{Naive}} = 0$.

\subsection*{Simplified three-node model}
\label{subsec:three_node_model}

\begin{figure}[H]
    \centering
    \includegraphics[width=1\linewidth]{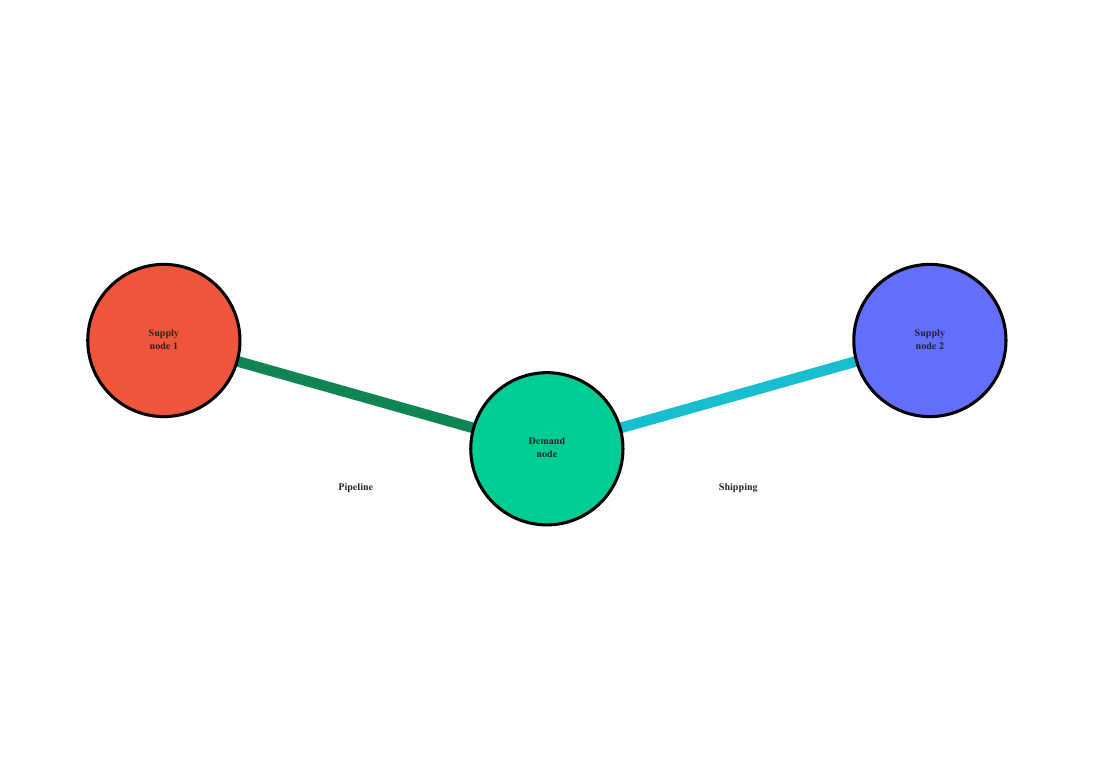}
    \caption{\textbf{Graph of the three-node model.} The figure shows the three nodes, one demand node and two supply nodes and the different characteristics of the edges. In green a pipeline connection to supply node one and in blue a shipping connection to supply node two.}
    \label{fig:3_node_toy}
\end{figure}

We analytically solve a simplified version of the model that contains only one demand node and two potential supply nodes and two realisations (Fig.~\ref{fig:3_node_toy}). The first supply node exhibits lower supply costs but its infrastructure entails risks (i.e. a pipeline connection), while the second supply node is assumed free of risks but more expensive (i.e. imports of a liquid hydrogen carrier like ammonia). In this simplified version of the model, the generation and transport capacities are not restricted, and the nodes can generate unlimited quantities of hydrogen.

When omitting the constraints on maximal capacity and maximal generation, the full model can be reduced to Karusch-Kuhn-Tucker (KKT) form with Lagrange multipliers $\la1^r_{n,n'}$ and $\la2_{n,n'}^r$ and slackness condition $\sa2_{n,n'}^r$ to yields (see Supplementary Information):
\begin{align}
    0 ~=~ & c^f_{n,n'} - \sum_{r^*} \la2_{n,n'}^{r^*} af_{n,n'}^{r^*} \\
    0 ~=~ & p^r \cdot cg_n - \la1_n^r \\
    0 ~=~ & p^r \cdot m_n \cdot [\Upsilon_n^r - d_n] + \la1_n^r \\
    0 ~=~ & p^r \cdot c^t_{n,n'} + \la1_n^r - \la1_{n'}^r + \la2_{n,n'}^r \\
    0 ~=~ & \Gamma_n^r - \Upsilon_n^r + \sum_{\mathclap{n'\neq n}} \left( \Phi_{n',n}^r - \Phi_{n,n'}^r \right) \\
    0 ~=~ & K_{n,n'} \cdot af_{n,n'}^r - \Phi_{n,n'}^r - \sa2_{n,n'}^r \punc.
\end{align}

We then make use of the simplifying assumption that there are only one demand and two supply nodes and two realisations with probability $p^{(1)} = 1-p$ and $p^{(2)} = p$. This allows us to find the capacities that minimise $TC$:
\begin{align}
    K_{n,n'}^{\min} = d_{n'} - \frac{\frac{c^f_{n,n'}}{p^r} + c^t_{n,n'} + c^g_n}{m_{n'}} \punc.
\end{align}
By omitting $n' = 3$ as an index, we find
\begin{align}
    K_n^{\min} = d - \alpha_n \punc,
\end{align}
where we defined $\alpha_n = \frac{\frac{c^f_n}{p^n} + c^t_n + c^g_n}{m}$. Alternatively, when the connection is not built, it is $K_n = 0$ (and therefore $\Phi_n^r = \Gamma_n^r = 0 ~~ \forall r\in R$).

We now consider two cases: Case I, where only the cheap connection is built, and Case II, where both connections are built. We can calculate $TC$ for the two cases when with the respective optimal capacity being built, which we define as $TC_I$ and $TC_{II}$.
\begin{align}
    TC_I &= \frac{m}{2} \cdot \left(2 \, d \, \alpha_1 - [\alpha_1]^2 \right) + p \frac{m}{2} \left[ d - \alpha_1 \right]^2 \\
    TC_{II} &= \frac{m}{2} \cdot \left(2 \, d \, \alpha_1 - [\alpha_1]^2 \right) + p \frac{m}{2} \left( 2 \, d \, \alpha_2 - [\alpha_2]^2 - 2 \, d \, \alpha_1 + [\alpha_1]^2 \right)
\end{align}

This allows us to compute
\begin{equation}
    \frac{W_{\mathrm{Anticipating}}-W_{\mathrm{Naive}}}{W_{\mathrm{Idealised}}} = p \cdot \left(\frac{m \,d  - \frac{c^f_2}{1-p} - c^t_2 - c^g_2}{m\,d - c^f_1 - c^t_1 - c^g_1}\right)^2\punc,
\label{eq:rho}
\end{equation}
which for large $m \cdot d$ can expanded to the expression given in the main text. For $p>0$, Eq. \eqref{eq:rho} is only zero when
\begin{align}
    p = \frac{c^f_2}{m\,d - c^t_2 - c^g_2} \punc,
\end{align}
which we define as  $p_0$.

\subsection*{Empirical Parameters}
\label{sec:parameters}

For the empirical study the model is parametrised for Europe in 2050. We group the member states of the European Union into five demand nodes. As we focus on net imports to Europe, all European nodes have a maximum hydrogen generation capacity of zero, while the generation capacity of the supply nodes is based on \cite{Genge2025}. First, Central Europe (CEE) that includes the Netherlands, Belgium, Luxembourg, Germany, Switzerland, Austria, Poland, and the Czech Republic. Second, South-West Europe (SWE) including France, Spain, and Portugal. Third, South-Central Europe (SCE) representing Italy, Slovenia, and Croatia. Fourth, South-East Europe (SEE) that includes the Slovak Republic, Hungary, Romania, Bulgaria, and Greece. And fifth, North-East Europe (NEE) - Denmark, Sweden, Finland, Estonia, Latvia, and Lithuania. Our model does not include Cyprus and Malta, as these countries are not connected to natural gas pipelines. This clustering mainly follows the European Hydrogen Backbone initiative \cite{Guidehouse2022}. There is one difference to be noted, Denmark is added to the north Europe node, representing the strong connection between the Danish and Swedish gas market. Our model includes three supply nodes for gaseous hydrogen. First, North-West Europe (NWE) representing Norway, Great Britain, and Ireland. Ireland is included in this node, as there are no direct connections from Ireland to Europe mainland. This node has the potential to be connected to Central Europe and West Europe nodes (CEE \& SWE). Second, West-North Africa (MeNa West) (MNW) that includes Morocco and Algeria. This node can be connected to the South-West Europe node (SWE). And third, East-North Africa (MeNa East) (MNE) representing Tunisia and Libya with a possible connection to the South-Central Europe node (SCE). \\
Today, the vast majority of energy carriers are imported via pipeline or ship \cite{Khan2019}. Therefore, we consider edges to be either pipelines or shipping connections. The possible connections between the nodes mentioned before are based on existing natural gas pipelines. Finally, we assume one node for ammonia as a liquid hydrogen carrier (NH3). This node can be connected to all demand nodes through shipping.\\
A pipeline's fixed and variable transport costs differ depending on the length and kind (submarine or onshore) of the respective connection. Variable transport costs include all transport related costs e.g. operation of compressor stations or conversion and re-conversion costs. 
Fixed investment and variable transport costs for shipping connections are the same for all edges. Variable generation costs are the sum of hydrogen generation costs at the production node. For the generation costs, we assume a global plateau at 80~EUR~\si{\per\MWh}  
and calculate the supply potential of generation nodes following \cite{Genge2025}. \\
We assume a completed transformation to a net zero energy system based on renewable energy sources. This includes a complete switch from natural gas to hydrogen. We therefore exclude natural gas as a fall-back option and assume an existing distribution grid for hydrogen within the nodes. For this reason, and due to the strong clustering, the resulting network is shorter than, for example, the assumed network of the European Hydrogen Backbone Initiative~\cite{Guidehouse2022}. To correct for the reduction of the network within the clustering, we corrected the pipeline distances within Europe with the factor 1.5~\cite{Brown2018,Frysztacki2021}. We also tested different distance correction factors to validate our results. Fig.~\ref{fig:extended data} presents the welfare effects for the used factor 1.5 as well as factors 5 and 10. The figure highlights, that the welfare gains from anticipatory planing are robust While the resulting optimal investments change.
We assume an equal demand response of 5~EUR~\si{\per\TWh\squared}  
for all nodes. We are aware that this is significantly higher than \cite{Weienburger2024} found. However, we use a short-term demand response, while \cite{Weienburger2024} addresses a long-term demand response. Therefore, we assume that industries that rely on hydrogen can switch on the long-run but are dependent on a hydrogen supply in the short-run and thus have a higher willingness to pay. \\

The probabilities for outages for the uncertain import connections are calculated following eq.~\ref{eq:risk}. Where $p_c$ is the probability of outages, $i_{base}$ is the interest rate for 'certain' countries in central Europe based on \cite{Moritz2023}, $i_c$ is the interest rate for a country also based on \cite{Moritz2023}. As each supply node contains at least two countries, we use the average for the nodes: $p_n = \frac{p_{c1} + p_{c2}}{2}$. As this method only allows for a very rough estimate of the default risks, the results are rounded to whole percentages.\\
This approach neglects that an investor requests an additional risk premium for an unsecure investment \cite{Damodaran2003}. We assume a risk-neutral investor who requires the same expected return from a certain and an uncertain investment. Thus, different interest rates $i_n$ reflect the expected default risk. The probabilities for outages ($p$) are  8~\% for connections from P1 (No \& Gb), 37~\% for the connection from P2 (Ma \& Dz), and 41~\% for the connection from P3 (Tn \& Ly), this is calculated based on eq.~\ref{eq:risk}.

\begin{equation}
   p_c = 1-\frac{i_{base}}{i_c}
    \label{eq:risk}
\end{equation}

\begin{figure}[H]
    \centering
    \begin{subfigure}{.7\textwidth}
        \includegraphics[width=1\linewidth]{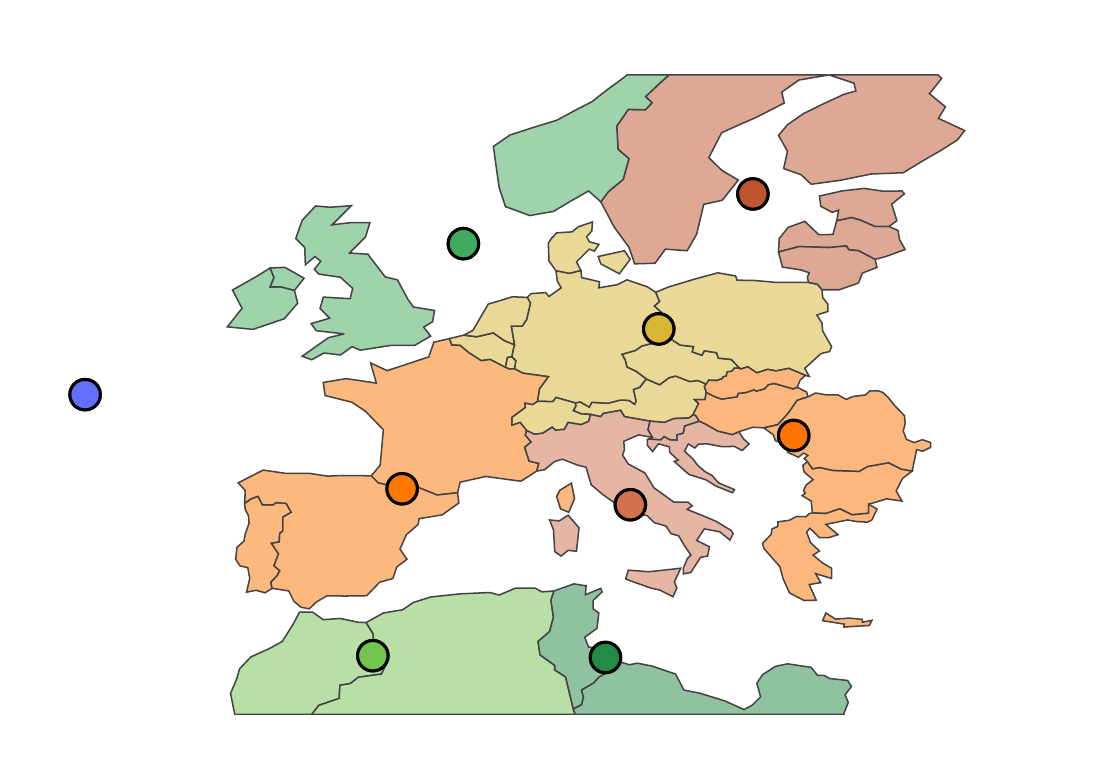}
    \end{subfigure}
    \begin{subfigure}{.7\textwidth}
        \includegraphics[width=1\linewidth]{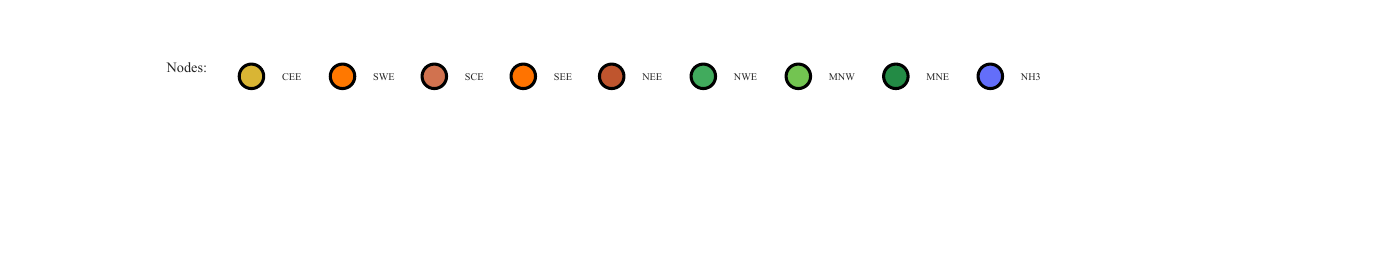}
    \end{subfigure}
    \caption{\textbf{Geographical resolution of the study} The European countries clustered to five demand nodes in  brown / orange. Surrounded by four supply nodes. Green are supply nodes that can supply hydrogen via pipelines and blue is the supply node for ammonia.}
    \label{fig:nodes}
\end{figure}

Fig~\ref{fig:nodes} shows the nodes.

\newpage

\subsubsection*{Availability of Connections}
\label{subsec:emirical_availability}

\begin{figure}[H]
    \centering
    \includegraphics[width=1\linewidth]{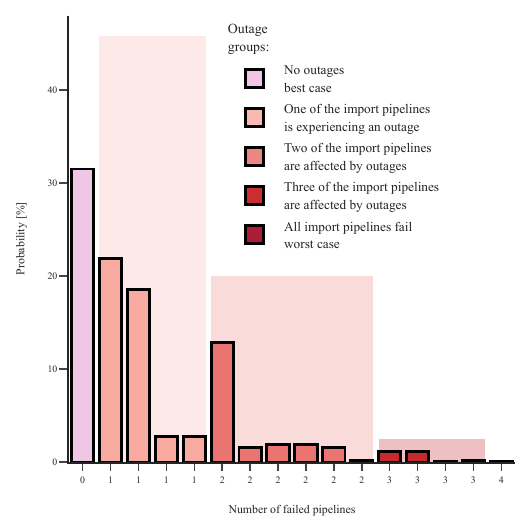}
    \caption{\textbf{Probabilities of realisations.}  The probability distribution of all outage realisations is shown, with each of the 16 realisations represented as a bar. The realisations are grouped by the number of failed pipelines and the probability of any realisation in a group to occur is shown as a lighter shape behind each group of bar. It shows that, while the likeliest realisation is the one without outages at \SI{30}{\percent}, the likeliest group of realisations is the one with exactly one failed pipeline at \SI{45}{\percent}. }
    \label{fig:realisations}
\end{figure}

\begin{figure}[H]
    \centering
    \includegraphics[width=.8\linewidth]{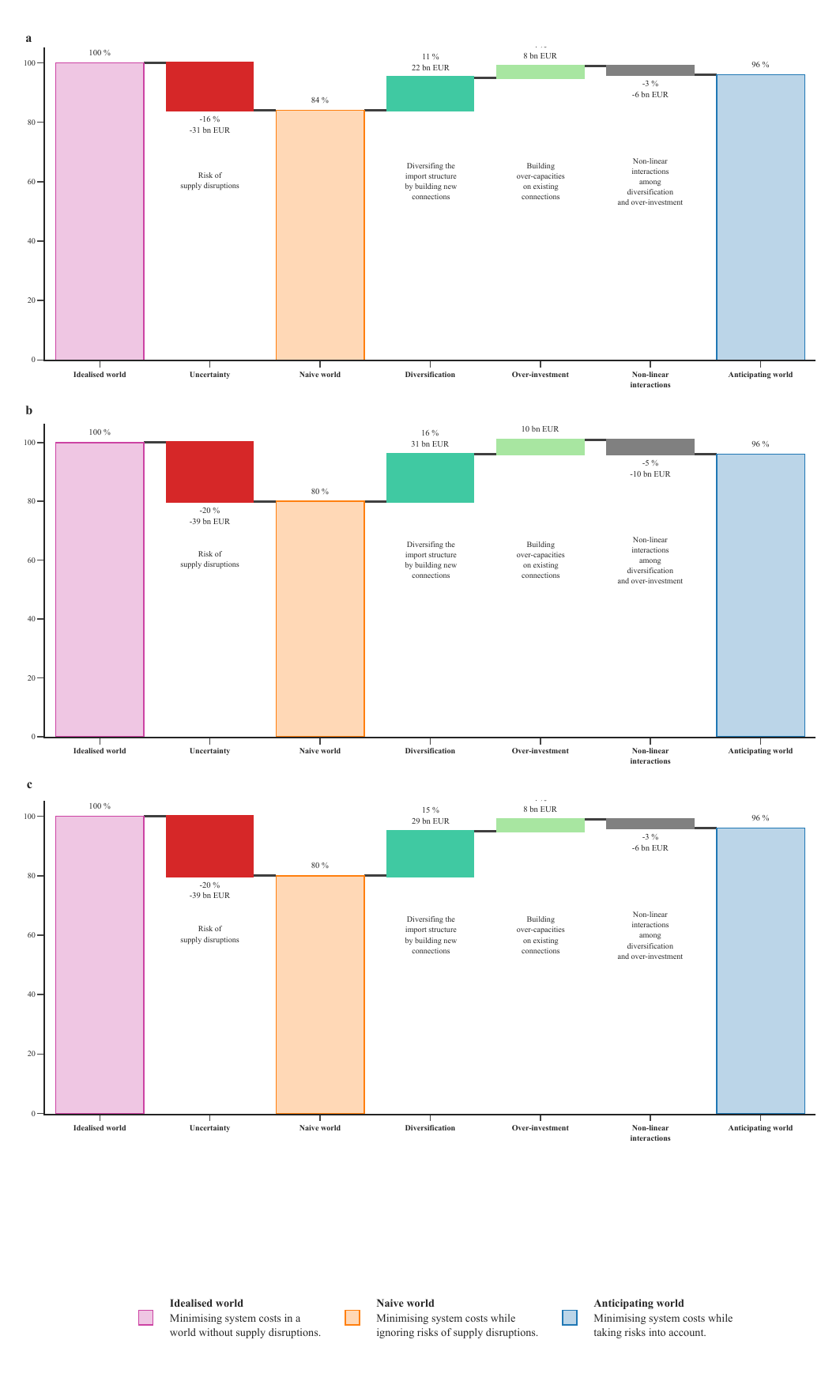}
    \caption{\textbf{Welfare effects under different distance correction factors.} We compare three correction factors: 1.5, 5, and 10. Compared to the literature these are relatively high, but consider the significant clustering used in our study. \textbf{a}, Factor 1.5,  \textbf{b}, Factor 5, and  \textbf{c}, Factor 10. The figures highlight that the overall welfare effect is robust to these correction factors.}
    \label{fig:extended data}
\end{figure}

In our model, risk occurs in the form of pipeline outages. As we assume individual and independent chances for outages for each pipeline import edge, there are 16 different risk realisations. These represent different combinations of functional and unusual pipeline edges. The probabilities for pipeline failures are presented in table~\ref{tab:probabilities} and Fig.~\ref{fig:realisations}. Table~\ref{tab:availabilities_empirical} shows the availabilities in all realisations. Fig.~\ref{fig:realisations} shows the realisations grouped by the number of failed pipeline edges. The pipelines are named according to the scheme ['Origin node'-'Destination node']

\begin{threeparttable}
    \caption{Edge assumptions}
    \label{tab:probabilities}
    \begin{tabular}{ccccc} \toprule
    Level        & {[}NWE-CEE{]} & {[}NWE-SWE{]} & {[}MNW-SWE{]} & {[}MNE-SCE{]} \\
                & \%          & \%          & \%          & \%           \\ \midrule
                   &                      &                      &                      &                      \\ 
    very   low     & 0                    & 0                    & 5                    & 10                   \\
    lower          & 2.5                  & 2.5                  & 10                   & 15                   \\
    low            & 5                    & 5                    & 15                   & 20                   \\
    mid-low        & 6.5                  & 6.5                  & 25                   & 30                   \\
    medium\tnote{†} & 8                   & 8                    & 37                   & 41                   \\
    mid-high       & 15                   & 15                   & 42                   & 50                   \\
    high           & 20                   & 20                   & 45                   & 60                   \\
    higher         & 25                   & 25                   & 52                   & 70                   \\
    very   high    & 30                   & 30                   & 60                   & 80                   \\  
                   &                      &                      &                      &                      \\ \bottomrule
    \end{tabular}
    \begin{tablenotes}\footnotesize
        \item[†] Uncertainty level used for the empirical study
    \end{tablenotes}
\end{threeparttable}

\begin{table}[h]
    \caption{Uncertainty realisations and availability factors}
    \label{tab:availabilities_empirical}
    \centering
    \begin{tabular}{lcccccc} \toprule 
         Realisation & \multicolumn{5}{c}{Availability factors for pipelines and shipping} & Probability  \\
               & [NWE-CEE] & [NWE-SWE] & [MNW-SWE] & [MNE-SCE] &  Shipping  &               \\  
             &       &       &       &      &       & \%        \\ \midrule 
             &       &       &       &      &       &           \\
1            & 1     & 1     & 1     & 1    & 1     & 31.46     \\
2            & 1     & 1     & 0     & 0    & 1     & 12.84     \\
3            & 1     & 1     & 1     & 0    & 1     & 21.86     \\
4            & 1     & 1     & 0     & 1    & 1     & 18.48     \\
5            & 0     & 1     & 1     & 1    & 1     & 2.74      \\
6            & 0     & 1     & 0     & 1    & 1     & 1.61      \\
7            & 0     & 1     & 1     & 0    & 1     & 1.9       \\
8            & 0     & 1     & 0     & 0    & 1     & 1.12      \\
9            & 1     & 0     & 1     & 1    & 1     & 2.74      \\
10           & 1     & 0     & 0     & 0    & 1     & 1.12      \\
11           & 1     & 0     & 1     & 0    & 1     & 1.9       \\
12           & 1     & 0     & 0     & 1    & 1     & 1.61      \\
13           & 0     & 0     & 1     & 1    & 1     & 0.24      \\
14           & 0     & 0     & 0     & 1    & 1     & 0.14      \\
15           & 0     & 0     & 1     & 0    & 1     & 0.17      \\
16           & 0     & 0     & 0     & 0    & 1     & 0.1       \\
             &       &       &       &      &       &           \\ \bottomrule       
    \end{tabular}
\end{table}
\clearpage

\subsubsection*{Pipeline and terminal assumptions}
\label{subsec:pipe_and_termina_assumptions}

Our assumptions for the pipelines from North Africa are based on \cite{Timmerberg2019}, the assumptions for the Norway pipelines are based on \cite{Fritz2009}, and the assumptions for the UK pipelines are based on \cite{Futyan2006}. Detailed information is listed in the supplementary data.xlsx file. Table~\ref{tab:pipes_final} shows the used data for pipeline edges.
The parameters used for the terminal connections are shown in table~\ref{tab:terminal} and are based on \cite{Ishimoto2020}.

\begin{threeparttable}
    \caption{Edge assumptions}
    \label{tab:pipes_final}
    \begin{tabular}{lccccc} \toprule 
    Pipeline            & Capacity  & Investment costs & Fixed O\&M    & Transport costs     & Probability \\
    
                        & \si{\TWh}  & $\frac{Million~EUR}{\si{\TWh}}$   & $\frac{Million~EUR}{\si{\TWh}}$ & $\frac{Million~EUR}{\si{\TWh}}$       & \% \\ \midrule 
                        
                        &           &           &           &       &       \\
    NWE-CEE             & 781       & 1.82      & 0.46      & 0.64  &  8    \\
    NWE-SWE             & 159       & 2.74      & 0.69      & 0.96  &  8    \\
    MNW-SWE             & 160       & 1.82      & 0.26      & 0.52  & 20    \\
    MNE-SCE             & 330       & 2.78      & 0.34      & 0.60  & 30    \\
                        &           &           &           &       \\
    \bottomrule           
    \end{tabular}
\end{threeparttable}

\begin{threeparttable}
    \caption{Terminal parameters}
\label{tab:terminal}
\begin{tabular}{lc|c} 
Total   investment            & Million EUR                 & 294        \\
Capacity - NH$_3$             & t~\si{\per\hour}              & 120        \\
Hydrogen   recovery factor    &                              & 0.695      \\
Hydrogen load factor          & $\si{\MWh}_{H_2}$~${t_{NH_3}}^{-1}$             & 5.8245     \\
Capacity - H$_2$              & Million~\si{\euros\per\TWh} & 4.25 \\
Levelised costs               & Million~\si{\euros\per\TWh} & 6.2  \\
\end{tabular}
\end{threeparttable}

\renewcommand*{\thefootnote}{\arabic{footnote}}

\subsubsection*{General techno-economic assumptions}
\label{subsec:general_assumptions}
Table~\ref{tab:parameters} presents the general techno-economic parameters. All energy units correspond to the lower heating value of hydrogen. A transport capacity of one~\si{\TWh} within one year eq. $\frac{1}{8.76}~\si{\mega\watt}$ or $\frac{1}{0.365}~\si{\GWh\per\day}$.

\begin{table}[h]
\caption{Parameters}
\footnotesize
\label{tab:parameters}
\setcounter{mpFootnoteValueSaver}{\value{footnote}} \centering
\begin{tabular}{lccc} \toprule 
Parameter                    & Unit                         & Value            & Source\\ \midrule 
                             &                              &                  &  \\
Time   horizon               & Year                         & one (2050)       &  \\ 
                             &                              &                  &  \\
Ship   transport             &                              &                  &  \\
Conversion and reconversion  & \si{\euros\per\MWh}    &  45              & \cite{Genge2023}\cite{Genge2025} \\
Terminal   investment        & Million \si{\euros\per\TWh} &      6.2    & \cite{Ishimoto2020} \\
Cost recovery factor         &   -                          &      0.0817      & \cite{Ishimoto2020} \\
                             &                              &                  &  \\
Pipeline   transport         &                              &                  &  \\
Refurbishment   – onshore    & \si{\euros\per\mega\watt\per\kilo\meter} &    129.4682      & \cite{Wang2021} \\ 
Lifetime - onshore pipeline  & years                        & 50               & \cite{DEA2021} \\
FOM - onshore pipeline       & \% p.a.                      & 1.5              & \cite{DEA2021} \\
Refurbishment   – offshore   & \si{\euros\per\mega\watt\per\kilo\meter} &    160.1566      & \cite{Wang2021} \\ 
Lifetime - offshore pipeline & years                        & 30               & \cite{dAmore-Domenech2021} \\
FOM - offshore pipeline      & \% p.a.                      & 3                & \cite{Fasihi2017} \\
Variable   transport costs   & (\si{\euros\per\MWh} \& 1000~\si{\kilo\meter})   &    0.51          &  \cite{Wang2020,ACER2021}\\
Pipeline   length – onshore  & \si{\kilo\meter}             & various values \footnotemark                & Own definition \\
Pipeline   length – offshore & \si{\kilo\meter}             &  various values \footnotemark & \cite{Timmerberg2019,SeaCorridor_2023,greenstream,MEDGAZ,Naturgy2016,entsog_2024} \\ 
Maximum pipeline capacity    & \si{\TWh}         &  various values \footnotemark & \cite{Timmerberg2019,ACER2021,Sagdur2023,entsog_2024} \\
Weighted average cost of capital (WACC) &                    &   8~\% \footnotemark          & \cite{Brändle2021, Wolf2024} \\

                             &                              &                  &  \\
Hydrogen   generation costs  & \si{\euros\per\MWh}          & 80               & \cite{Genge2023,Genge2025} \\
                             &                              &                  &  \\
Pipelines                    &                              &                  &  \\
Chance for   outages         & \%                           & various values \footnotemark  & \cite{Moritz2023,Damodaran2025} \\
                             &                              &                  &  \\
Market                       &                              &                  &  \\
Latent  import demand        & \si{\TWh}                    & 505~\footnotemark & \cite{RePowerEU, ENTSOG2025} \\
Demand   price response      & Million \si{\euros\per\TWh\squared} & 5 & Own estimation \\
                             &                              &                  & \\ \bottomrule 
\end{tabular}%
\stepcounter{mpFootnoteValueSaver}%
    \footnotetext[\thempFootnoteValueSaver]{%
        100 km in exporting country to coast and distance coast to import node 'middle'}%
\stepcounter{mpFootnoteValueSaver}
\footnotetext[\thempFootnoteValueSaver]{%
    We use the average length of the existing natural gas pipelines 
}
\stepcounter{mpFootnoteValueSaver}
\footnotetext[\thempFootnoteValueSaver]{%
    We assume refurbishment of existing natural gas pipelines allowing 80~\% of the original pipeline capacity being available for hydrogen transport
}
\stepcounter{mpFootnoteValueSaver}
\footnotetext[\thempFootnoteValueSaver]{We assume the 8~\% WACC as base value for European countries and add a country risk premium following \cite{Damodaran2003,Damodaran2022,Damodaran2025} for supply nodes}
\stepcounter{mpFootnoteValueSaver}
\footnotetext[\thempFootnoteValueSaver]{Based on equation~\ref{eq:risk}, we calculate the default risk based on interest rates for countries (given by \cite{Moritz2023}) and country risk premiums (given by \cite{Damodaran2025})}
\stepcounter{mpFootnoteValueSaver}
\footnotetext[\thempFootnoteValueSaver]{We use the net imports assumed in \cite{ENTSOG2025} and correct them with an assumed price of 63~EUR per MWh \cite{Koirala2021} to get a latent demand per node.}
\end{table}

\subsubsection*{Parameterisation of simplified model}
\label{subsec:toy_availability}

Fixed investment costs as well as variable transport costs of the uncertain option are significant lower compared to the reliable option \cite{Ishimoto2020,Timmerberg2019,Fritz2009}. In this simplified version both connections' capacities are not restricted and the nodes can generate unlimited quantities of hydrogen. Tab.~\ref{tab:toy_params} shows the chosen parameters.

\begin{table}[h]
    \caption{Three-node model parameters}
    \label{tab:toy_params}
    \centering
    \begin{tabular}{lllllll} \toprule
    $m_n$ & $d_n$ & $c^f_1$ & $c^f_2$ & $c^t_1$ & $c^t_2$ & $c^g_{1,2}$ \\ 
      &     &     &     &     &     &    \\ \midrule
    5 & 500 & 3 & 6  & 0.5   & 45   & 80 \\ \bottomrule
    \end{tabular}
\end{table}


\section*{Data availability} Input and output data are available as a ZIP archive. A full overview of files, formats, and their interrelationships is given in file `README.md`.

\section*{Code availability} The code is available as a ZIP archive containing Python and GAMS source code files. A full overview of files, formats, and their interrelationships is given in file `README.md`.

\clearpage
\bibliography{export}

\begin{thebibliography}{10}

\bibitem{RePowerEU}
European Commission.
\newblock Implementing the repower eu action plan: Investment needs, hydrogen accelerator and achieving the bio-methane targets.
\newblock Technical report, European Commission, 2022.

\bibitem{Niermann2021}
M.~Niermann, S.~Timmerberg, S.~Drünert, and M.~Kaltschmitt.
\newblock Liquid organic hydrogen carriers and alternatives for international transport of renewable hydrogen.
\newblock {\em Renewable and Sustainable Energy Reviews}, 135:110171, 1 2021.

\bibitem{ACER2021}
ACER.
\newblock Transporting pure hydrogen by repurposing existing gas infrastructure: Overview of existing studies and reflections on the conditions for repurposing.
\newblock Technical report, European Union Agency for the Cooperation of Energy Regulators, 2021.

\bibitem{Lee2014}
Jay~H. Lee.
\newblock Energy supply planning and supply chain optimization under uncertainty.
\newblock {\em Journal of Process Control}, 24:323--331, 2 2014.

\bibitem{Moritz2023}
Michael Moritz, Max Schönfisch, and Simon Schulte.
\newblock Estimating global production and supply costs for green hydrogen and hydrogen-based green energy commodities.
\newblock {\em International Journal of Hydrogen Energy}, 48:9139--9154, 3 2023.

\bibitem{Sens2022_1}
Lucas Sens, Ulf Neuling, Karsten Wilbrand, and Martin Kaltschmitt.
\newblock Conditioned hydrogen for a green hydrogen supply for heavy duty-vehicles in 2030 and 2050 – a techno-economic well-to-tank assessment of various supply chains.
\newblock {\em International Journal of Hydrogen Energy}, 1 2022.

\bibitem{Sens2022_2}
Lucas Sens, Yannick Piguel, Ulf Neuling, Sebastian Timmerberg, Karsten Wilbrand, and Martin Kaltschmitt.
\newblock Cost minimized hydrogen from solar and wind – production and supply in the european catchment area.
\newblock {\em Energy Conversion and Management}, 265, 8 2022.

\bibitem{franzmann2023}
D.~Franzmann, H.~Heinrichs, F.~Lippkau, T.~Addanki, C.~Winkler, P.~Buchenberg, T.~Hamacher, M.~Blesl, J.~Linßen, and D.~Stolten.
\newblock Green hydrogen cost-potentials for global trade.
\newblock {\em International Journal of Hydrogen Energy}, 48:33062--33076, 10 2023.

\bibitem{benalcazar2024}
P.~Benalcazar and A.~Komorowska.
\newblock Techno-economic analysis and uncertainty assessment of green hydrogen production in future exporting countries.
\newblock {\em Renewable and Sustainable Energy Reviews}, 199:114512, 7 2024.

\bibitem{Brandt2024}
Jonathan Brandt, Thore Iversen, Christoph Eckert, Florian Peterssen, Boris Bensmann, Astrid Bensmann, Michael Beer, Hartmut Weyer, and Richard Hanke-Rauschenbach.
\newblock Cost and competitiveness of green hydrogen and the effects of the european union regulatory framework.
\newblock {\em Nature Energy}, 9:703--713, 5 2024.

\bibitem{Genge2023}
Lucien Genge, Fabian Scheller, and Felix Müsgens.
\newblock Supply costs of green chemical energy carriers at the european border: A meta-analysis.
\newblock {\em International Journal of Hydrogen Energy}, 48:38766--38781, 12 2023.

\bibitem{Genge2025}
Lucien Genge and Felix Müsgens.
\newblock Green ammonia: A techno-economic supply chain optimization.
\newblock 7 2025.

\bibitem{Genge2026}
Lucien Genge, Marius Neuwirth, Khaled Al-Dabbas, Tobias Fleiter, and Felix Müsgens.
\newblock Optimising green value chains for the chemical industry in europe.
\newblock {\em International Journal of Hydrogen Energy}, 199:152689, 1 2026.

\bibitem{Riera2023}
Jefferson~A. Riera, Ricardo~M. Lima, and Omar~M. Knio.
\newblock A review of hydrogen production and supply chain modeling and optimization.
\newblock {\em International Journal of Hydrogen Energy}, 48:13731--13755, 4 2023.

\bibitem{Odenweller2022}
Adrian Odenweller, Falko Ueckerdt, Gregory~F. Nemet, Miha Jensterle, and Gunnar Luderer.
\newblock Probabilistic feasibility space of scaling up green hydrogen supply.
\newblock {\em Nature Energy}, 7:854--865, 9 2022.

\bibitem{Egli2025}
Florian Egli, Flurina Schneider, Alycia Leonard, Claire Halloran, Nicolas Salmon, Tobias Schmidt, and Stephanie Hirmer.
\newblock Mapping the cost competitiveness of african green hydrogen imports to europe.
\newblock {\em Nature Energy}, 10:750--761, 6 2025.

\bibitem{Aldren2025}
Cameron Aldren, Nilay Shah, and Adam Hawkes.
\newblock Quantifying key economic uncertainties in the cost of trading green hydrogen.
\newblock {\em Cell Reports Sustainability}, 2:100342, 5 2025.

\bibitem{Kim2024}
Sunwoo Kim, Joungho Park, Wonsuk Chung, Derrick Adams, and Jay~H. Lee.
\newblock Techno-economic analysis for design and management of international green hydrogen supply chain under uncertainty: An integrated temporal planning approach.
\newblock {\em Energy Conversion and Management}, 301:118010, 2 2024.

\bibitem{Rezaee2024}
A.~Rezaee Jordehi, Seyed~Amir Mansouri, Marcos Tostado-Véliz, Miguel Carrión, M.~J. Hossain, and Francisco Jurado.
\newblock A risk-averse two-stage stochastic model for optimal participation of hydrogen fuel stations in electricity markets.
\newblock {\em International Journal of Hydrogen Energy}, 49:188--201, 1 2024.

\bibitem{Almansoori2012}
A.~Almansoori and N.~Shah.
\newblock Design and operation of a stochastic hydrogen supply chain network under demand uncertainty.
\newblock {\em International Journal of Hydrogen Energy}, 37:3965--3977, 3 2012.

\bibitem{Caglayan2021}
Dilara~Gulcin Caglayan, Heidi~U. Heinrichs, Martin Robinius, and Detlef Stolten.
\newblock Robust design of a future 100
\newblock {\em International Journal of Hydrogen Energy}, 46:29376--29390, 8 2021.

\bibitem{Yokoyama2018}
Ryohei Yokoyama, Akira Tokunaga, and Tetsuya Wakui.
\newblock Robust optimal design of energy supply systems under uncertain energy demands based on a mixed-integer linear model.
\newblock {\em Energy}, 153:159--169, 6 2018.

\bibitem{Emenike2020}
Scholastica~N. Emenike and Gioia Falcone.
\newblock A review on energy supply chain resilience through optimization.
\newblock {\em Renewable and Sustainable Energy Reviews}, 134:110088, 12 2020.

\bibitem{Lambert2022}
Laurent~A. Lambert, Jad Tayah, Caroline Lee-Schmid, Monged Abdalla, Ismail Abdallah, Abdalftah~H.M. Ali, Suhail Esmail, and Waleed Ahmed.
\newblock The eu's natural gas cold war and diversification challenges.
\newblock {\em Energy Strategy Reviews}, 43:100934, 9 2022.

\bibitem{DeRosa_2022}
Mattia~De Rosa, Kenneth Gainsford, Fabiano Pallonetto, and Donal~P. Finn.
\newblock Diversification, concentration and renewability of the energy supply in the european union.
\newblock {\em Energy}, 253:124097, 8 2022.

\bibitem{Dantzig1955}
George~B. Dantzig.
\newblock Linear programming under uncertainty.
\newblock {\em Management Science}, 1:197--206, 4 1955.

\bibitem{Riepin2021}
Iegor Riepin, Thomas Möbius, and Felix Müsgens.
\newblock Modelling uncertainty in coupled electricity and gas systems—is it worth the effort?
\newblock {\em Applied Energy}, 285:116363, 3 2021.

\bibitem{Antweiler2025}
Werner Antweiler and Felix Muesgens.
\newblock The new merit order: The viability of energy-only electricity markets with only intermittent renewable energy sources and grid-scale storage.
\newblock {\em Energy Economics}, 145:108439, 5 2025.

\bibitem{Kotek2023}
Peter Kotek, Borbála~Takácsné Tóth, and Adrienn Selei.
\newblock Designing a future-proof gas and hydrogen infrastructure for europe – a modelling-based approach.
\newblock {\em Energy Policy}, 180:113641, 9 2023.

\bibitem{Kountouris2024}
Ioannis Kountouris, Rasmus Bramstoft, Theis Madsen, Juan Gea-Bermúdez, Marie Münster, and Dogan Keles.
\newblock A unified european hydrogen infrastructure planning to support the rapid scale-up of hydrogen production.
\newblock {\em Nature Communications}, 15:5517, 6 2024.

\bibitem{Verpoort2024}
Philipp~C. Verpoort, Lukas Gast, Anke Hofmann, and Falko Ueckerdt.
\newblock Impact of global heterogeneity of renewable energy supply on heavy industrial production and green value chains.
\newblock {\em Nature Energy}, 9:491--503, 4 2024.

\bibitem{Odenweller2025}
Adrian Odenweller and Falko Ueckerdt.
\newblock The green hydrogen ambition and implementation gap.
\newblock {\em Nature Energy}, 10:110--123, 1 2025.

\bibitem{herbst2012}
Andrea Herbst, Felipe Toro, Felix Reitze, and Eberhard Jochem.
\newblock Introduction to energy systems modelling.
\newblock {\em Swiss Journal of Economics and Statistics}, 148:111--135, 4 2012.

\bibitem{pfenninger2014}
Stefan Pfenninger, Adam Hawkes, and James Keirstead.
\newblock Energy systems modeling for twenty-first century energy challenges.
\newblock {\em Renewable and Sustainable Energy Reviews}, 33:74--86, 5 2014.

\bibitem{Riepin2019}
Iegor Riepin and Felix Müsgens.
\newblock Seasonal flexibility in the european natural gas market.
\newblock Technical report, 2019.

\bibitem{bernecker2022}
Maximilian Bernecker, Iegor Riepin, and Felix Müsgens.
\newblock Modeling of extreme weather events—towards resilient transmission expansion planning.
\newblock In {\em 2022 18th International Conference on the European Energy Market (EEM)}, pages 1--7. IEEE, 9 2022.

\bibitem{DeJonghe2012}
Cedric~De Jonghe, Benjamin~F. Hobbs, and Ronnie Belmans.
\newblock Optimal generation mix with short-term demand response and wind penetration.
\newblock {\em IEEE Transactions on Power Systems}, 27:830--839, 5 2012.

\bibitem{Antweiler2021}
Werner Antweiler and Felix Muesgens.
\newblock On the long-term merit order effect of renewable energies.
\newblock {\em Energy Economics}, 99, 7 2021.

\bibitem{Guidehouse2022}
Guidehouse.
\newblock Five hydrogen supply corridors for europe in 2030.
\newblock Technical report, European Hydrogen Backbone Initiative, 5 2022.

\bibitem{Khan2019}
Nasrullah Khan, Saad Dilshad, Rashida Khalid, Ali~Raza Kalair, and Naeem Abas.
\newblock Review of energy storage and transportation of energy.
\newblock {\em Energy Storage}, 1, 6 2019.

\bibitem{Brown2018}
T.~Brown, D.~Schlachtberger, A.~Kies, S.~Schramm, and M.~Greiner.
\newblock Synergies of sector coupling and transmission reinforcement in a cost-optimised, highly renewable european energy system.
\newblock {\em Energy}, 160:720--739, 10 2018.

\bibitem{Frysztacki2021}
Martha~Maria Frysztacki, Jonas Hörsch, Veit Hagenmeyer, and Tom Brown.
\newblock The strong effect of network resolution on electricity system models with high shares of wind and solar.
\newblock {\em Applied Energy}, 291:116726, 6 2021.

\bibitem{Weienburger2024}
Bastian Weißenburger, Martin Wietschel, Benjamin Lux, and Matthias Rehfeldt.
\newblock The long term price elastic demand of hydrogen – a multi-model analysis for germany.
\newblock {\em Energy Strategy Reviews}, 54:101432, 7 2024.

\bibitem{Damodaran2003}
Aswath Damodaran.
\newblock Country risk and company exposure: Theory and practice.
\newblock {\em Journal of Applied Finance}, 13:63, 10 2003.

\bibitem{Timmerberg2019}
Sebastian Timmerberg and Martin Kaltschmitt.
\newblock Hydrogen from renewables: Supply from north africa to central europe as blend in existing pipelines – potentials and costs.
\newblock {\em Applied Energy}, 237:795--809, 3 2019.

\bibitem{Fritz2009}
Petr Fritz, Hakan Sköldberg, and Bo~Rydén.
\newblock Final report - nordic energy perspectives natural gas in the nordic countries.
\newblock Technical report, Nordic Energy perspectives, 2009.

\bibitem{Futyan2006}
Mark. Futyan.
\newblock {\em The Interconnector Pipeline - A key link in Europe's Gas Network}.
\newblock Oxford Institute for Energy Studies, 2006.

\bibitem{Ishimoto2020}
Yuki Ishimoto, Mari Voldsund, Petter Nekså, Simon Roussanaly, David Berstad, and Stefania~Osk Gardarsdottir.
\newblock Large-scale production and transport of hydrogen from norway to europe and japan: Value chain analysis and comparison of liquid hydrogen and ammonia as energy carriers.
\newblock {\em International Journal of Hydrogen Energy}, 45:32865--32883, 11 2020.

\bibitem{Wang2021}
Anthony Wang, Jaro Jens, David Mavins, Marissa Moultak, Matthias Schimmel, Kees van~der Leun, Daan Peters, and Maud Buseman.
\newblock European hydrogen backbone - analysing future demand, supply, and transport of hydrogen.
\newblock Technical report, European Hydrogen Backbone, 2021.

\bibitem{DEA2021}
Energy transport - technology descriptions and projections for long-term energy system planning.
\newblock Technical report, The Danish Energy Agency, 2021.

\bibitem{dAmore-Domenech2021}
Rafael d'Amore Domenech, Teresa~J. Leo, and Bruno~G. Pollet.
\newblock Bulk power transmission at sea: Life cycle cost comparison of electricity and hydrogen as energy vectors.
\newblock {\em Applied Energy}, 288:116625, 4 2021.

\bibitem{Fasihi2017}
Mahdi Fasihi, Dmitrii Bogdanov, and Christian Breyer.
\newblock Long-term hydrocarbon trade options for the maghreb region and europe-renewable energy based synthetic fuels for a net zero emissions world.
\newblock 2017.

\bibitem{Wang2020}
Anthony Wang, Kees van~der Leun, Daan Peters, and Maud Buseman.
\newblock European hydrogen backbone how a dedicated hydrogen infrastructure can be created.
\newblock Technical report, Guidehouse, 2020.

\bibitem{SeaCorridor_2023}
SeaCorridor.
\newblock How to become a client of ttpc and transmed in a couple of steps.
\newblock Technical report, SeaCorridor, 2023.

\bibitem{greenstream}
greenstream bv.
\newblock Greenstream pipeline, 2024.

\bibitem{MEDGAZ}
MEDGAZ S.A.
\newblock Technical summary, 2024.

\bibitem{Naturgy2016}
Naturgy.
\newblock 20th anniversary of the maghreb–europe gas pipeline, 12 2016.

\bibitem{entsog_2024}
ENTSOG.
\newblock System capacity map 2024.
\newblock Technical report, ENTSOG, 2024.

\bibitem{Sagdur2023}
Yasin Sagdur, Robert Slowinski, Rik van Rossum, Alex Kozub, Luis Kühnen, Martijn Overgaag, Alizé Michelet, Stepahanie Kandathiparambil, and Poppy London.
\newblock Implementation roadmap-cross border projects and costs update european hydrogen backbone.
\newblock Technical report, Guidehouse, 11 2023.

\bibitem{Brändle2021}
Gregor Brändle, Max Schönfisch, and Simon Schulte.
\newblock Estimating long-term global supply costs for low-carbon hydrogen.
\newblock {\em Applied Energy}, 302:117481, 11 2021.

\bibitem{Wolf2024}
Nicolas Wolf, Michelle~Antje Tanneberger, and Michael Höck.
\newblock Levelized cost of hydrogen production in northern africa and europe in 2050: A monte carlo simulation for germany, norway, spain, algeria, morocco, and egypt.
\newblock {\em International Journal of Hydrogen Energy}, 69:184--194, 6 2024.

\bibitem{Damodaran2025}
Aswath Damodaran.
\newblock Country default spreads and risk premiums, 1 2025.

\bibitem{ENTSOG2025}
ENTSOG and ENTSO-E.
\newblock Tyndp 2024 - scenarios report.
\newblock Technical report, ENTSOG, ENTSO-E, 2025.

\bibitem{Damodaran2022}
Aswath Damodaran.
\newblock Country risk: Determinants, measures and implications - the 2022 edition.
\newblock 7 2022.

\bibitem{Koirala2021}
Binod Koirala, Sebastiaan Hers, Germán Morales-España, Özge Özdemir, Jos Sijm, and Marcel Weeda.
\newblock Integrated electricity, hydrogen and methane system modelling framework: Application to the dutch infrastructure outlook 2050.
\newblock {\em Applied Energy}, 289:116713, 5 2021.

\end{thebibliography}


\section*{Funding}
This research is supported by the Federal Ministry of Research, Technology and Space (ARIADNE Award No. 03SFK5O0-2) as well as the German Federal Government, the Federal Ministry of Research, Technology and Space, and the State of Brandenburg within the framework of the joint project EIZ: Energy Innovation Center (project numbers 85056897 and 03SF0693A) with funds from the Structural Development Act (Strukturstärkungsgesetz) for coal-mining regions, and the HyValue project (grant no. 333151).

\section*{Author contribution}
S.M.R. and F.M. designed the study and developed the optimisation model. S.M.R. performed the modelling and analysis and drafted the paper. P.C.V. contributed to plotting the figures and drafting the paper. P.C.V. provided the analytical solution of the simplified model and contributed to the final draft. F.M. directed and supervised this project. S.M.R., P.C.V., F.U., and F.M. discussed the results and commented on the paper.

\section*{Ethics declarations}

\paragraph{Declaration of interests} F.M. is an employee and shareholder of r2b energy consulting GmbH.

\paragraph{Competing interests} The authors declare no competing interests.

\paragraph{Consent for publication} Not applicable.

\section*{Supplementary information}

The Supplementary Information (attached at the end of this file) provides a step-by-step derivation of the analytical solution of the simplified three-node model.

\newpage

\section*{Extended data figure}


\newpage
\section*{Supplementary information}
The following sections derive the analytical solution of the simplified three-node problem discussed in the main article.

\subsection*{Prerequisites}
Let $\ntot\in\mathbb{N}, \ntot\geq3$ be the total number of nodes, and let $N = \{1,\ldots,\ntot\}$ be the set of nodes. Let $\rtot\in\mathbb{N}$ be the total number of realisations, and let $R = \{1,\ldots,\rtot\}$ be the set of realisations.
$\forall n,n'\in N, r\in R$ let $m_n, c^f_{n,n'}, c^t_{n,n'}, c^g_{n}, d_n\in\RRgz$ and $af^r_{n,n'}\in\{0,1\}$ and $p^r\in(0,1]$. Further, we assume $\forall n,n'\in N, n\neq n'~\exists r\in R: af^r_{n,n'} = 1$.

Let $V = V_K \times V_\Gamma \times V_\Upsilon \times V_\Phi$ be the full variable space, where $V_K = \RRgz^{N(N-1)}$, ~ $V_\Gamma = \RRgz^{N R}$, ~ $V_\Upsilon = \RRgz^{N R}$, ~ $V_\Phi = \RRgz^{N(N-1)R}$. Let $K \in V_K$, $\Gamma \in V_\Gamma$, $\Upsilon \in V_\Upsilon$, and $\Phi \in V_\Phi$ be denoted as respectively
\begin{align}
  K &= (K_{n,n'})_{\substack{n,n'\in N\\n\neq n'}} &&= (K_{1, 2}, K_{1, 3}, ~\ldots~, K_{\ntot, \ntot-1}) \notag\\
  \Gamma &= (\Gamma_n^r)_{\substack{n\in N\\r\in R}} &&= (\Gamma_1^1, \Gamma_2^1, ~\ldots~, \Gamma_{\ntot}^{\rtot}) \notag\\
  \Upsilon &= (\Upsilon_n^r)_{\substack{n\in N\\r\in R}} &&= (\Upsilon_1^1, \Upsilon_2^1, ~\ldots~, \Upsilon_{\ntot}^{\rtot}) \notag\\
  \Phi &= (\Phi_{n,n'}^r)_{\substack{n,n'\in N\\r\in R\\n\neq n'}} &&= (\Phi_{1, 2}^1, \Phi_{1, 3}^1, ~\ldots~, \Phi_{\ntot, \ntot-1}^{\rtot}) \notag
\end{align}

Let the total cost function be defined as
\begin{align}
    & TC: V \rightarrow \mathbb{R} \\
\intertext{and}
    & TC(K, \Gamma, \Upsilon, \Phi) \\
=~~ & \sum_{\mathclap{\substack{n,n'\in N\\n\neq n'}}} K_{n,n'} \cdot c^f_{n,n'} \notag\\
+   & \sum_{\mathclap{r\in R}} p^r \sum_{\mathclap{n\in N}} \Bigg( \Gamma_n^r \cdot c^g_n + \sum_{\mathclap{\substack{n'\in N\\n'\neq n}}} \Phi_{n,n'}^r \cdot c^t_{n,n'} \notag\\
    & ~~~~~~~~~ + \frac{m_n}{2} \Bigg[\Upsilon_n^r - d_n\Bigg]^2 \Bigg) \punc .\notag
\end{align}

For $n\in N,r\in R$ let the first constraint functions $\Co1_n^r$ be defined as
\begin{align}
    & \Co1_n^r: V \rightarrow \mathbb{R}
\intertext{and}
    & \Co1_n^r(K, \Gamma, \Upsilon, \Phi) \\
=~~ & \Gamma_n^r - \Upsilon_n^r + \sum_{\mathclap{n'\neq n}} \left( \Phi_{n',n}^r - \Phi_{n,n'}^r \right) \punc. \notag
\end{align}

For $n,n'\in N,n\neq n',r\in R$ let the second constraint functions $\Co2_{n,n'}^r$ be defined as
\begin{align}
    & \Co2_{n,n'}^r: V \rightarrow \mathbb{R}
\intertext{and}
    & \Co2_{n,n'}^r (K, \Gamma, \Upsilon, \Phi) \\
=~~ & K_{n,n'} \cdot af_{n,n'}^r - \Phi_{n,n'}^r \punc. \notag
\end{align}

\subsection*{Problem}
We want to solve the following problem. Find $(K, \Gamma, \Upsilon, \Phi) \in V$ such that
\begin{align}
    \qquad \min                \quad & TC (K, \Gamma, \Upsilon, \Phi) \label{eq:problem_orig}\\[0.4cm]
           \text{subject to}   \quad & \Co1_n^r (K, \Gamma, \Upsilon, \Phi) = 0 \\
                                     & \forall n\in N, \forall r\in R \notag \\[0.4cm]
                                     & \Co2_{n,n'}^r (K, \Gamma, \Upsilon, \Phi) \geq 0 \\
                                     & \forall n,n'\in N, n\neq n', \forall r\in R \notag \\[0.4cm]
                                     & K_{n,n'} \geq 0 ~ \forall n,n'\in N, n\neq n' \\
                                     & \Gamma_n^r \geq 0 ~ \forall n\in N, \forall r\in R \\
                                     & \Upsilon_n^r \geq 0 ~ \forall n\in N, \forall r\in R \\
                                     & \Phi_{n,n'}^r \geq 0 ~ \forall n,n'\in N, n\neq n', \forall r\in R
\end{align}

\subsection*{KKT form}
This problem can be brought into the Krush-Kuhn-Tucker (KKT) form with complementary slackness conditions. Specifically, $(K, \Gamma, \Upsilon, \Phi)\in V$ is a solution to the problem defined in \eqnref{problem_orig} exactly when $\sa2\in\RRgz^{N(N-1)R}$ (so-called slack variables) such that $(K, \Gamma, \Upsilon, \Phi)\in V$ is a solution to the following problem.
\begin{align}
    \qquad \min                \qquad & TC(K, \Gamma, \Upsilon, \Phi) \label{eq:problem_kkt_start}\\[0.4cm]
           \text{subject to}   \qquad & 0 = \Co1_n^r (K, \Gamma, \Upsilon, \Phi) \\
                                      & \forall n\in N, \forall r\in R \notag \\[0.4cm]
                                      & 0 = \Co2_{n,n'}^r (K, \Gamma, \Upsilon, \Phi) - \sa2_{n,n'}^r \\
                                      & \forall n,n'\in N, n\neq n', \forall r\in R \notag \\[0.4cm]
                                      & K_{n,n'} \geq 0 ~ \forall n,n'\in N, n\neq n' \\
                                      & \Gamma_n^r \geq 0 ~ \forall n\in N, \forall r\in R \\
                                      & \Upsilon_n^r \geq 0 ~ \forall n\in N, \forall r\in R \\
                                      & \Phi_{n,n'}^r \geq 0 ~ \forall n,n'\in N, n\neq n', \forall r\in R \label{eq:problem_kkt_end}
\end{align}

\subsection*{Euler-Lagrange-Equations}
The problem defined in Eqns.~\eqref{eq:problem_kkt_start}--\eqref{eq:problem_kkt_end} can be solved with Lagrange multipliers $\la1\in\mathbb{R}^{NR}$ and $\la2\in\mathbb{R}^{N(N-1)R}$. First, we define the Lagrange function $\Lagrange$:
\begin{align}
    & \Lagrange(K, \Gamma, \Upsilon, \Phi, \la1, \la2, \sa2) \notag \\[0.4cm]
= ~~ & TC(K, \Gamma, \Upsilon, \Phi) \notag \\[0.4cm]
  -  & ~\sum_{\mathclap{n^*,r^*}}~ \la1_{n^*}^{r^*} ~ \Co1_{n^*}^{r^*}(K, \Gamma, \Upsilon, \Phi) \notag \\[0.4cm]
  -  & ~\sum_{\mathclap{\substack{n^*,n^{**},r^*\\n^*\neq n^{**}}}}~ \la2_{n^*,n^{**}}^{r^*} \left( \Co2_{n^*,n^{**}}^{r^*}(K, \Gamma, \Upsilon, \Phi) - \sa2_{n^*,n^{**}}^{r^*} \right) \notag
\end{align}
The Euler-Lagrange-Equation
\begin{align}
  0 = & \quad \nabla_{K, \Gamma, \Upsilon, \Phi, \la1, \la2} ~ \Lagrange
\intertext{then yields}
  0 = & \quad \pdb{TC}{K_{n,n'}} \\
      & - ~\sum_{\mathclap{n^*,r^*}}~ \la1_{n^*}^{r^*} ~ \pdb{\Co1_{n^*}^{r^*}}{K_{n,n'}}  \notag\\
      & - ~\sum_{\mathclap{n^*,n^{**}}^{r^*}}~ \la2_{n^*,n^{**}}^{r^*} ~ \pdb{\Co2_{n^*,n^{**}}^{r^*}}{K_{n,n'}}  \notag\\
      & \quad \forall n,n'\in N,n\neq n' \punc, \notag\\[0.7cm]
  0 = & \quad \pdb{TC}{\Gamma_n^r} \\
      & - ~\sum_{\mathclap{n^*,r^*}}~ \la1_{n^*}^{r^*} ~ \pdb{\Co1_{n^*}^{r^*}}{\Gamma_n^r}  \notag\\
      & - ~\sum_{\mathclap{n^*,n^{**},r^*}}~ \la2_{n^*,n^{**}}^{r^*} ~ \pdb{\Co2_{n^*,n^{**}}^{r^*}}{\Gamma_n^r}  \notag\\
      & \quad \forall n\in N, \forall r\in R \punc, \notag\\[0.7cm]
  0 = & \quad \pdb{TC}{\Upsilon_n^r} \\
      & - ~\sum_{\mathclap{n^*,r^*}}~ \la1_{n^*}^{r^*} ~ \pdb{\Co1_{n^*}^{r^*}}{\Upsilon_n^r}  \notag\\
      & - ~\sum_{\mathclap{n^*,n^{**},r^*}}~ \la2_{n^*,n^{**}}^{r^*} ~ \pdb{\Co2_{n^*,n^{**}}^{r^*}}{\Upsilon_n^r}  \notag\\
      & \quad \forall n\in N, \forall r\in R \punc, \notag\\[0.7cm]
  0 = & \quad \pdb{TC}{\Phi_{n,n'}^r} \\
      & - ~\sum_{\mathclap{n^*,r^*}}~ \la1_{n^*}^{r^*} ~ \pdb{\Co1_{n^*}^{r^*}}{\Phi_{n,n'}^r}  \notag\\
      & - ~\sum_{\mathclap{n^*,n^{**},r^*}}~ \la2_{n^*,n^{**}}^{r^*} ~ \pdb{\Co2_{n^*,n^{**}}^{r^*}}{\Phi_{n,n'}^r}  \notag\\
      & \quad \forall n,n'\in N,n\neq n', \forall r\in R \punc, \notag\\[0.7cm]
  0 = & \quad \Co1_n^r (K, \Gamma, \Upsilon, \Phi) \\
      & \quad \forall n\in N, \forall r\in R \punc, \notag\\
  0 = & \quad \Co2_{n,n'}^r (K, \Gamma, \Upsilon, \Phi) - \sa1_{n,n'}^r \\
      & \quad \forall n,n'\in N,n\neq n', r\in R \punc, \notag \\
\end{align}
while
\begin{align}
      & K_{n,n'} \geq 0 ~ \forall n,n'\in N, n\neq n' \\
      & \Gamma_n^r \geq 0 ~ \forall n\in N, \forall r\in R \\
      & \Upsilon_n^r \geq 0 ~ \forall n\in N, \forall r\in R \\
      & \Phi_{n,n'}^r \geq 0 ~ \forall n,n'\in N, n\neq n', \forall r\in R
\end{align}
\\

Specifically, this yields
\begin{align}
    0 ~=~ & cf_{n,n'} - \sum_{r^*} \la2_{n,n'}^{r^*} af_{n,n'}^{r^*} \\
    0 ~=~ & p^r \cdot cg_n - \la1_n^r \\
    0 ~=~ & p^r \cdot m_n \cdot [\Upsilon_n^r - d_n] + \la1_n^r \\
    0 ~=~ & p^r \cdot c^t_{n,n'} + \la1_n^r - \la1_{n'}^r + \la2_{n,n'}^r \\
    0 ~=~ & \Gamma_n^r - \Upsilon_n^r + \sum_{\mathclap{n'\neq n}} \left( \Phi_{n',n}^r - \Phi_{n,n'}^r \right) \\
    0 ~=~ & K_{n,n'} \cdot af_{n,n'}^r - \Phi_{n,n'}^r - \sa2_{n,n'}^r \punc.
\end{align}

\subsection*{Simplification}
The simplification mentioned in Methods is achieved by assuming $\ntot = 3$, $\Upsilon_n^r = 0$ for $n\in\{1, 2\}$, $\Gamma_n^r = 0$ for $n=3$, and $\Phi_{n,n'}^r = 0$ for $n\in\{1,2\}$ and $n'=3$. Moreover, let's assume that $\rtot = 2$ and that $af_{1,3}^r = 0$ for $r=2$.

We now simplify this problem, by assuming that there are only 1 demand and 2 supply nodes and two realisations with probability $p^{(1)} = 1-p$ and $p^{(2)} = p$ (Fig. 6a). This allows us to find the capacities that minimise $TC$:
\begin{align}
    K_{n,n'}^{\min} = d_{n'} - \frac{\frac{c^f_{n,n'}}{p^r} + c^t_{n,n'} + c^g_n}{m_{n'}} \punc.
\end{align}
By omitting $n' = 3$ as an index, we find
\begin{align}
    K_n^{\min} = d - \alpha_n \punc,
\end{align}
where we defined $\alpha_n = \frac{\frac{c^f_n}{p^n} + c^t_n + c^g_n}{m}$. Alternatively, when the connection is not built, it is $K_n = 0$ (and therefore $\Phi_n^r = \Gamma_n^r = 0 ~~ \forall r\in R$).

We now consider two cases: Case I, where only the cheap connection is built, and Case II, where both connections are built. We can calculate $TC$ for the two cases when with the respective optimal capacity being built, which we define as $TC_I$ and $TC_{II}$.
\begin{align}
    TC_I = ~ & (1-p) \cdot K_1^{\min} \cdot \underbrace{\left(\frac{c^f_1}{1-p} + c^g_1 + c^t_1 \right)}_{= m \cdot \alpha_1} \\
             & +~ (1-p) \frac{m}{2} \underbrace{\left[ K_1^{\min} - d \right]^2}_{=[\alpha_1]^2} \notag \\
             & +~ p \frac{m}{2} \left[ d \right]^2 \notag \punc, \\
         = ~ & (d - \alpha_1) \cdot m \cdot \alpha_1 + \frac{m}{2} \cdot [\alpha_1]^2 \\
             & + p \left( \frac{m}{2} [d]^2 - (d - \alpha_1) \cdot m \cdot \alpha_1 - \frac{m}{2} [\alpha_1]^2 \right) \notag \\
         = ~ & \frac{m}{2} \cdot \left(2 \, d \, \alpha_1 - [\alpha_1]^2 \right) \\
             & + p \frac{m}{2} \left[ d - \alpha_1 \right]^2 \notag \\
    TC_{II} = ~ & (1-p) \cdot K_1^{\min} \cdot \underbrace{\left(\frac{c^f_1}{1-p} + c^g_1 + c^t_1 \right)}_{= m \alpha_1} \\
             & + p \cdot K_2^{\min} \cdot \underbrace{\left(\frac{c^f_2}{p} + c^g_2 + c^t_2 \right)}_{= m \alpha_2} \notag \\
             & +~ (1-p) \frac{m}{2} \underbrace{\left[ K_1^{\min} - d \right]^2}_{=[\alpha_1]^2} \notag \\
             & +~ p \frac{m}{2} \underbrace{\left[ K_2^{\min} - d \right]^2}_{=[\alpha_2]^2} \punc, \notag \\
         = ~ & (d - \alpha_1) \cdot m \cdot \alpha_1 + \frac{m}{2} \cdot [\alpha_1]^2 \\
             & + p \bigg( \frac{m}{2} [\alpha_2]^2 + (d - \alpha_2) \cdot m \cdot \alpha_2 \\
             & ~~~~ - \frac{m}{2} [\alpha_1]^2 - (d - \alpha_1) \cdot m \cdot \alpha_1 \bigg) \notag \\
         = ~ & \frac{m}{2} \cdot \left(2 \, d \, \alpha_1 - [\alpha_1]^2 \right) \\
             & + p \frac{m}{2} \left( 2 \, d \, \alpha_2 - [\alpha_2]^2 - 2 \, d \, \alpha_1 + [\alpha_1]^2 \right) \notag
\end{align}

We now aim to compute
\begin{equation}
    \rho = \frac{W_{\mathrm{Anticipating}}-W_{\mathrm{Naive}}}{W_{\mathrm{Idealised}}} \punc.
\end{equation}
These components can be computed from $W_{\mathrm{Anticipating}}-W_{\mathrm{Naive}} = TC_I - TC_{II}$ and $W_{\mathrm{Idealised}} = W_\max - TC_0$, where $TC_0 = TC_I$ for $p=0$. Specifically, it is:
\begin{align}
    TC_{I} - TC_{II} &= p \cdot \frac{m \, \left(d - \alpha_2 \right)^2}{2} \\
    W_\max &= \frac{m\,d^2}{2} \\
    TC_0   &= \frac{m \, \left(2 \, d \, \alpha_0 - [\alpha_0]^2 \right)}{2}
\intertext{with $\alpha_0 = \frac{c^f_1 + c^t_1 + c^g_1}{m}$, and therefore}
    \rho &= p \left(\frac{d-\alpha_2}{d-\alpha_0}\right)^2 = p \left(\frac{m\,d - \frac{c^f_2}{1-p} - c^g_2 - c^t_2}{m\,d - c^f_1 - c^g_1 - c^t_1}\right)^2.
\end{align}

Since $\rho = 0$ when $TC_{I} - TC_{II} = 0$, we can use this to define as $p_0$ for which $\rho = 0$:
\begin{align}
    p_0 = \frac{c^f_2}{m\,d - c^t_2 - c^g_2} \punc.
\end{align}

\end{document}